\documentclass[prb,aps,amsmath,amssymb,reprint,showpacs,floatfix,superscriptaddress]{revtex4-1}
\usepackage{graphicx}
\usepackage{dcolumn}
\usepackage{bm}
\usepackage{mathtools}
\usepackage{enumitem}

\begin{document}


\title{Skyrmion spin transfer torque due to current confined in a nanowire} 



\author{Javier Osca}
\email{javier.osca@imec.be}
\affiliation{IMEC, Kapeldreef 75, B-3001 Leuven, Belgium}
\affiliation{KU Leuven, ESAT-MICAS, Kasteelpark Arenberg 10, 3001 Leuven, Belgium}
\author{Bart Sor\'ee}
\affiliation{IMEC, Kapeldreef 75, B-3001 Leuven, Belgium}
\affiliation{KU Leuven, ESAT-MICAS, Kasteelpark Arenberg 10, 3001 Leuven, Belgium}
\affiliation{Universiteit Antwerpen, Departement Fysica, B-2000 Antwerpen, Belgium}

\date{May 17, 2020}

\begin{abstract}
In this work we compute the torque field present in a ferromagnet in contact with a metallic nanowire when a skyrmion  is present. If the nanowire is narrow enough the current is carried by a single conduction band. In this regime the classical torque model breaks down and we show that a skyrmion driven by spin transfer torque moves in a different direction than predicted by the classical model. However, the amount of charge current required to move a skyrmion with a certain velocity in the single band regime is similar to a classical model of torque where it is implicitly assumed current transport by many conduction bands. The single band regime is more efficient creating spin current from charge current because of the perfect polarization of the single band but is less efficient creating torque from spin current. Nevertheless, it is possible to take profit of the single band regime to move skyrmions even with no net charge or spin current flowing between the device contacts. We have also been able to recover the classical limit considering an ensemble of only a few electronic states. In this limit we have discovered that electron diffusion needs to be considered even in ballistic nanowires due the effect of the skyrmion structure on the electron current. 
\end{abstract}


\maketitle 

\section{Introduction}
\label{S1}
Skyrmions where first introduced in 1962 by Tony Skyrme  as a nucleon model \cite{Skyrme}. They have since found application in condensed matter physics within the field of spintronics. In this context, a skyrmion is a topological magnetic structure characterized by a definite Chern number \cite{Topo}.
There are two kind of skyrmion structures, Bloch skyrmions characterized by azimuthal magnetization in the skyrmion boundary and N\'eel skyrmions where the magnetization is radial on the boundary \cite{Fert,Tomasello}. Skyrmions are created by the balance between exchange and Dzyaloshinskii--Moriya interaction (DMI). Exchange interaction wants to drive the local magnetic moment of the ferromagnet to a minimum energy configuration where all the spins are aligned while the minimum energy configuration with DMI is attained where neighboring spins are perpendicular \cite{Dzyal,Moriya}. 

Bloch skyrmions have been found in ferromagnetic layers where the DMI originates from the inversion symmetry breaking within a unit cell of the crystal in combination of  the own material spin-orbit interaction \cite{Bloch1,Bloch2,Bloch3}. On the other hand, N\'eel skyrmions are found in interfaces between a ferromagnet and heavy metal layers where the symmetry breaking is caused by the interface \cite{Neel1,Neel2,Neel3}. A strong spin orbit coupling is also necessary which can be provided by the heavy metal. In this work we will focus on this last case because it will provide to us a larger control of the device physical geometry.

Skyrmion movement is possible using spin transfer torque (STT) or spin orbit torque (SOT) mechanisms \cite{RevTorques}. Both mechanisms have a common origin in the sd exchange interaction between the spin of the conduction electrons and the local magnetization of the ferromagnet. Torque models driven by spin currents may be applied to domain walls and other magnetic structures\cite{Torque} and in presence of skyrmions this torque induces skyrmion movement \cite{Fert,Tomasello}. Skyrmion movement by spin torque has been proposed as basis of logic \cite{Processor1,Processor2} and memory \cite{Tomasello,Fert,Kang} devices. The breakdown of Moore's law for small devices in CMOS technology has led to an increasing interest of spintronic technologies with skyrmions because the limits in power consumption and stability of future CMOS and memory devices. 

Skyrmion memory devices\cite{Tomasello} were proposed as an improvement over the racetrack memory\cite{IBM} where  the information is encoded in magnetic regions separated by domain walls (DW). In skyrmion racetrack memories the skyrmions take the role of the DW increasing the amount of information per unit surface and lowering the power consumption with respect their DW counterparts.

In this work we will focus in the movement of N\'eel magnetic skyrmions using spin transfer torque STT in a quasi-2d interface between a ferromagnet and a narrow metallic wire. A quantum mechanical approach for electron transport is needed to model conduction in a narrow nanowire where the transverse conduction bands are well separated in energy. Up to this moment most of the skyrmion movement by STT reported in literature \cite{STT1,STT2,STT3} has used the classical torque term proposed in Ref. \cite{Torque}. A few works have dealt with a quantum mechanical model of current in relation to skyrmion movement trough spin torque \cite{QTopological,QuantumFourier} but with a very different set of boundary conditions, resolution methods and objectives than this paper. 
 
In a narrow nanowire it is possible to achieve a sizable spin current due to the transport of electrons in a single fully polarized band. The resulting torque field from the interaction between the spin degrees of freedom of the conduction electrons and the skyrmion magnetic structure is different from the torque field predicted by the classical model\cite{Torque}. We will show how with spin current originating from a single polarized band it is possible to have skyrmion movement even in zero power conditions. Furthermore we will show how in this regime the presence of skyrmions leave a signature in the conductance of the nanowire current.

Finally, we will recover the classical torque field considering an ensemble of the spin of various electronic states in the multi-band regime. Comparing the result with the classical model for the torque we have discovered that the terms that arise due electron diffusion are needed even in ballistic nanowires due to the scattering effect of the magnetic skyrmion structure. 

The paper is divided in four sections:
{\bf Section \ref{S1}: Introduction} presenting the general concepts and background. {\bf Section \ref{S2}: Theoretical model and formalism} where it is described the numerical method needed to calculate the torque field and skyrmion movement. {\bf Section \ref{S3}: Results} where the topics summarized above are presented and discussed. {\bf Section \ref{S4}: Conclusions} presented as a summary of key results and some additional comments.

\section{Theoretical model and formalism}
\label{S2}
\subsection{Quantum model}
\begin{figure}
  \includegraphics[width=\columnwidth]{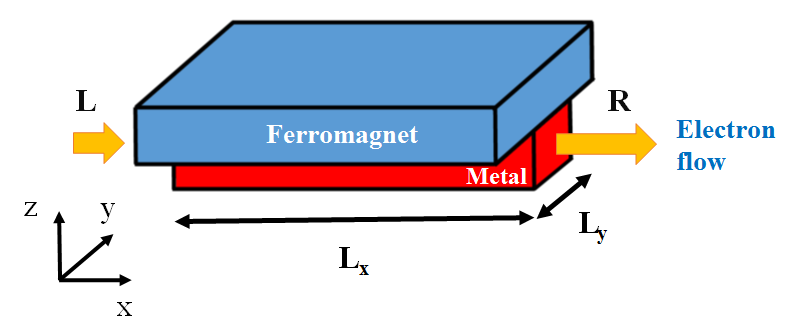}
  \caption{ Device model schematic. The skyrmion appears in the ferromagnet but is driven by the torque created by the conduction electrons of the nanowire.}
  \label{F1A}
\end{figure}

In this work we will consider a model of a quasi-2d interface between a ferromagnet and a metallic nanowire as shown fig.\,\ref{F1A}. The ferromagnet provides the non-itinerant spin degrees of freedom that support the skyrmion while an electronic current flows through the metallic nanowire connected to two terminals, left L and right R. The torque is caused by an action-reaction force applied in equal measure but with different sign on the skyrmion magnetic structure and on the itinerant electrons due the sd exchange interaction between the spin of the conduction electrons and the spin of the localized electrons in the ferromagnet. In order to model the dynamics of the magnetization we derive here the torque terms for the Landau-Lifshitz (LL) equation from a quantum mechanical model.

We start assuming that spins of itinerant and localized electrons interact at the interface where they are in close proximity to each other. The interface is modeled as a 2d grid where each position represents an atom of the crystal lattice separated by a distance $a_c$ . The interface between the ferromagnet and the nanowire is described by a quantum Hamiltonian,
\begin{equation}
\hat{\mathcal{H}}=\hat{\mathcal{H}}_{\rm 0}+\hat{\mathcal{H}}_{\rm dis}\,,
\label{E1}
\end{equation}
where  
\begin{equation}
\hat{\mathcal{H}}_{\rm 0}=\hat{\mathcal{H}}_{\rm Z}+\hat{\mathcal{H}}_{\rm ex}+\hat{\mathcal{H}}_{\rm DMI}+\hat{\mathcal{H}}_{\rm sd}+\hat{\mathcal{H}}_{\rm K}\,,
\label{E2}
\end{equation}
is the hermitian term of the hamiltonian that describes the different spin interactions present in the device where
$\hat{\mathcal{H}}_{\rm Z}$, $\hat{\mathcal{H}}_{\rm ex}$, $\hat{\mathcal{H}}_{\rm DMI}$, $\hat{\mathcal{H}}_{\rm sd}$ and $\hat{\mathcal{H}}_{\rm K}$ are the Hamiltonian terms
for the Zeeman interaction, exchange interaction, DMI, sd interaction and kinetic energy of the conduction electrons respectively. On the other hand, 
\begin{equation}
\hat{\mathcal{H}}_{\rm dis}=-i\lambda\hat{\mathcal{H}}_{\rm 0}\,,
\label{E3}
\end{equation}
is a non hermitian Hamiltonian that models energy dissipation as proposed in refs. \cite{Wieser1,Wieser2}.

In more detail, the exchange term models the spin-spin interaction between neighboring sites,
\begin{equation}
\hat{\mathcal{H}}_{\rm ex}=-\sum_i\sum_j  J_{i,j} \left( \bf{\hat{S}_i} \cdot \bf{\hat{S}_j} \right) \,,
\label{E5}
\end{equation}
where ${\bf \hat{S}_i}$ and ${\bf \hat{S}_j}$ are the dimensionless spin operators and $J_{i,j}=J$ if $i$ and $j$ are neighbors but zero otherwise.
The DMI term describes the antisymmetric exchange between two neighboring spins of the lattice,
\begin{equation}
\hat{\mathcal{H}}_{\rm DMI}=-\sum_i 	\sum_j {\bf D_{i,j}} \cdot \left(  {\bf \hat{S}_i}	\times {\bf \hat{S}_i} \right)  \,,
\label{E7}
\end{equation}
where ${\bf D_{i,j}}$ is different from zero when i and j are neighbors. The nanowire-ferromagnet interface breaks the inversion symmetry therefore only a perpendicular DMI term is present where ${\bf D_{i,j}}=D \left( {\bf \hat{z}} \times {\bf \hat{r}_{i,j}} \right)$ and where ${\bf \hat{r}_{i,j}}$ is the vector between two neighboring atomic sites. The balance between exchange and DMI forces is responsible for the creation of the skyrmion in the ferromagnet. An external Zeeman field or a magnetocrystalline anisotropy in the material is needed in order to stabilize the skyrmion \cite{Bogdanov,Size,Phases}. In this work we will consider without loss of generality an external Zeeman field for that purpose. 

The Zeeman term describes the interaction between an external magnetic field and the non-itinerant spins located at each of the atomic sites of the 2d crystal lattice, 
\begin{equation}
\hat{\mathcal{H}}_{\rm Z}=\gamma \hbar \sum_i \bf{\hat{S}_i} \cdot \bf{B}_i \,,
\label{E4}
\end{equation}
where $\gamma$ is the gyromagnetic factor and ${\bf B_i}$ is the magnetic field felt at the atom site $i$. On the other hand, $\hat{\mathcal{H}}_{\rm K}$ is the kinetic energy of the conduction electrons,
\begin{equation}
\hat{\mathcal{H}}_{\rm K}=\frac{{\bf \hat{p}}^2}{2m}\,,
\label{E7B}
\end{equation}
where ${\bf \hat{p}}=\sum_n {\bf \hat{p}}_n$, ${\bf \hat{p}}_n$ is the conduction electron $n$ momentum operator and  $m$ is the effective mass of the electrons. Note that itinerant electrons are not attached to any particular atomic site but are free to move around the whole wire therefore they are not labeled by any site index. 

Finally, the sd Hamiltonian term describes the interaction between the spin degrees of freedom of the conduction electrons and the spins pinned to atomic sites,
\begin{equation}
\hat{\mathcal{H}}_{\rm sd}=-J_{\rm sd}\sum_i {\bf\hat{s}_i} \cdot {\bf\hat{S}_i}\,,
\label{E7C}
\end{equation}
where in the same way as before ${\bf \hat{S}_i}$ is the dimensionless spin operator for the pinned electrons attached to the atomic site $i$.
On the other hand ${\bf \hat{s}_i}$ is the corresponding dimensionless spin operator for the conduction electrons contained in an atomic cell volume for the same atomic site. This is ${\bf\hat{s}_i}=V_{\rm c} \delta(r-r_i) {\bf \hat{s}}$ where $V_{\rm c}=a^3_{\rm c}$ and the dimensionless spin angular momentum operator is defined as ${\bf \hat{s}}=1/2(\hat{\sigma}_x,\hat{\sigma}_y,\hat{\sigma}_z)$ where $\hat{\sigma}_{x,y,z}$ are the corresponding Pauli matrices. This is the term that gives rise to the torque produced by spin currents.

\subsection{Generalized Ehrenfest theorem}
The non-Hermitian dissipation term in the Hamiltonian (eq.\,\ref{E1}) leads to a non conservation of the norm in time, for example,
\begin{equation}
n_l^2=\langle \tilde{\Psi}_l(t) | \tilde{\Psi}_l(t) \rangle=e^{-2\lambda\langle \tilde{\Psi}_l(t_0) | \hat{\mathcal{H}}_{\rm 0} | \tilde{\Psi}_l(t_0) \rangle  \Delta t } \,,
\label{E8}
\end{equation}
where $l$ is just a label for the Hamiltonian $\hat{\mathcal{H}}$ eigenstates $\tilde{\Psi}_l(t_0)$ and $\Delta_t=t-t_0$. According to refs. \cite{Wieser1,Wieser2} in order to enforce the conservation of the norm we can renormalize the eigenstates,
\begin{equation}
|\Psi_l(t) \rangle=\frac{| \tilde{\Psi}_l(t) \rangle }{\sqrt{1-r_l}}    \,,
\label{E9}
\end{equation}
where $r_l=-\frac{1}{i\hbar} \Delta t \langle \Psi_l(t_0) | \hat{\mathcal{H}}-\hat{\mathcal{H}}^\dagger |\Psi_l(t_0) \rangle$.  Note that $\tilde{\Psi}(t_0)=\Psi(t_0)$ at the initial time $t_0$. The time of evolution of the not normalized states is governed by, 
\begin{equation}
\frac{d|\tilde{\Psi}_l(t)\rangle}{dt} =\frac{1}{i\hbar}\hat{\mathcal{H}} |\tilde{\Psi}_l(t) \rangle   \,,
\label{E10}
\end{equation}
while for the normalized states,
\begin{equation}
\frac{d|\Psi_l(t_0)\rangle}{dt} =\frac{1}{i\hbar} \left( \hat{\mathcal{H}}_{\rm 0} -i\lambda \left(  \langle \Psi_l(t_0)| \hat{\mathcal{H}}_0 | \Psi_l(t_0) \rangle \right)\right) |\Psi_l(t_0) \rangle   \,,
\label{E11}
\end{equation}
gives as a result the usual Schr\"{o}dinger equation for an hermitian Hamiltonian plus a dissipation term. Although this equation is derived for $t_0$ the subindex can be dropped and eq.\,\ref{E11} can be applied to any time $t$ because the time origin $t_0$ is chosen arbitrarily. With this equation it is possible to obtain an expression for the expectation value of an operator analogous to the Ehrenfest theorem but applicable to an ensemble of mixed states and with an extra term accounting for dissipation, 
\begin{equation}
\begin{split}
\frac{d \langle\hat{O}\rangle}{dt}&=\frac{1}{i\hbar}\langle \left[\hat{O},\hat{\mathcal{H}}_{\rm 0} \right] \rangle\\
&-\frac{\lambda}{\hbar} \left( \langle \left\{  \hat{O},\hat{\mathcal{H}}_{\rm 0} \right\} \rangle \right.\\
&-2\left.\sum_l p_l \langle \Psi_l(t)|\hat{\mathcal{H}}_{\rm 0}|\Psi_l(t)\rangle  \langle \Psi_l(t)|\hat{O}|\Psi_l(t)\rangle\right)
\end{split}
\label{E12}
\end{equation}
where $\langle \hat{O}\rangle=Tr\left[\hat{\rho}\,\hat{O}\right]$ and $\hat{\rho}=\sum_l p_l |\Psi_l(t)\rangle \langle \Psi_l(t)|$. The last term in eq.\,\ref{E12} is different from the one proposed in \cite{Wieser2} because we are considering mixed states instead of pure ones. Nevertheless this difference becomes unimportant later on because for a ferromagnetic approximation the states of the ferromagnetic system are approximated to be locally pure.

With eq.\,\ref{E12} and the hermitian Hamiltonian $\mathcal{H}_{\rm 0}$ of eq.\,\ref{E2} we obtain the equation for the expectation value of the spin angular momentum at each of the atomic sites,
\begin{equation}
\begin{split}
\frac{d \langle {\bf \hat{S}_i} \rangle}{dt}=&-\gamma \left( {\bf B_i} \times \langle {\bf \hat{S}_i} \rangle \right)
+2\sum_j J_{i,j} \langle  {\bf \hat{S}_i}\times {\bf \hat{S}_j}  \rangle\\
&-2\sum_j {\bf D_{i,j}} \langle {\bf \hat{S}_i} \cdot {\bf \hat{S}_j} \rangle
+2\sum_j \langle \left({\bf D_{i,j}} \cdot {\bf \hat{S}_i} \right) {\bf \hat{S}_j} \rangle\\
&+J_{\rm sd}  \langle {\bf \hat{S}_i}\times {\bf \hat{s}_i} \rangle
-\frac{\lambda}{\hbar} \left( \langle \left\{  {\bf \hat{S}_i},\hat{\mathcal{H}}_{\rm 0} \right\} \rangle \right.\\
&-2\left.\sum_l p_l \langle \Psi_l(t)|\mathcal{H}_{\rm 0}|\Psi_l(t)\rangle  \langle \Psi_l(t)|{\bf \hat{S}_i}|\Psi_l(t)\rangle\right)
\end{split}\,.
\label{E13}
\end{equation}
Note that the Hamiltonian term corresponding to the kinetic energy of the conduction electrons $\hat{\mathcal{H}}_{\rm K}$ does not contribute to this equation because $[{\bf \hat{S}_i},{\bf \hat{S}_i}]=0$ and $[{\bf \hat{S}_i},{\bf \hat{s}_i}]=0$.

\subsection{Landau-Lifshitz equation derivation}
The Landau-Lifshitz equation is derived from eq.\,\ref{E13} under mean field approximation,
\begin{equation}
{\bf \hat{S}_i}\cdot{\bf \hat{S}_j}\approx {\bf \hat{S}_i}\langle{\bf \hat{S}_j}\rangle+\langle{\bf \hat{S}_i}\rangle {\bf \hat{S}_j} - \langle{\bf \hat{S}_i}\rangle \langle{\bf \hat{S}_j}\rangle\,,
\label{E14}
\end{equation}
neglecting second order terms. This approximation holds for ferromagnetic devices where the length of spatial variation of the pinned electrons spin is large enough in comparison with the atomic length $a_{\rm c}$. Furthermore, the dissipation term may be further simplified to the one depicted in ref. \cite{Wieser2} if we consider the equilibrium states to be locally pure because of the same slow spatial variation of the spins.

Once we have neglected second order correlation effects in eq.\,\ref{E13} we rewrite it into an equation of spin density. For each atomic site the spin density of the pinned electrons is ${\bf S(r_i)}=\rho_{\rm c} \langle {\bf \hat{S}_i} \rangle$ while  conduction electrons are treated on the same footing ${\bf s(r_i)}=\rho_{\rm c} \langle {\bf \hat{s}_i} \rangle$ where $\rho_{\rm c}=1/a_{\rm c}^3$.  In this form, we further approximate the spin density by a continuous field in the limit $a_{\rm c}=|{\bf r_i}-{\bf r_j}|\rightarrow 0$ where $i$ and $j$ are index for neighboring crystal sites. As a consequence we can apply the approximation, 
\begin{equation}
{\bf S(r_j)}\approx {\bf S(r_i)}+\left( {\bf r_{i,j}} \cdot \nabla \right) {\bf S(r_i)} + \frac{1}{2}\left( {\bf r_{i,j}}\cdot H \cdot {\bf r_{i,j}} \right) {\bf S(r_i)}
\end{equation}
where $H$ is the Hessian,
\begin{equation}
H=\begin{pmatrix}
  \frac{\partial}{\partial x^2}  & \frac{\partial}{\partial x\partial y}\\ 
  \frac{\partial}{\partial y\partial x} & \frac{\partial}{\partial y^2}
\end{pmatrix} \,.
\label{E15}
\end{equation}

Finally, the units are changed from angular momentum to magnetization ${\bf M(r_i)}=-\gamma {\bf S(r_i)}$, ${\bf m(r_i)}=-\gamma {\bf s(r_i)}$ where ${\bf M(r_i)}$ is the magnetization of pinned electrons at position ${\bf r_i}$ and ${\bf m(r_i)}$ is the magnetization of the conduction electrons on the same position. 

Applying the mean field approximation and the continuous spin field description we derive from eq.\,\ref{E13}: 
\begin{equation}
\begin{split}
\frac{d{\bf M(r)}}{dt}=&-\frac{\gamma_0}{1+\alpha^2} \left( {\bf M(r)} \times {\bf H}_{\rm eff}({\bf r}) \right)\\ &
-\frac{\alpha\gamma_0}{1+\alpha^2} \left({\bf M(r)} \times  \left( {\bf M(r)} \times {\bf H}_{\rm eff}({\bf r}) \right) \right) 
\end{split}
\label{E16}
\end{equation}
where $\gamma_0=\gamma\mu_0$, $\mu_0$ is the vacuum permeability and
\begin{equation}
{\bf H}_{\rm eff}({\bf r})={\bf H}_{\rm Z}({\bf r})+{\bf H}_{\rm ex}({\bf r})+{\bf H}_{\rm DMI}({\bf r})+{\bf H}_{\rm T}({\bf r})\,,
\label{E17}
\end{equation}
is the local effective magnetic field felt by the magnetization. There is one term for each of the interactions where the spin of the pinned electrons is involved. Those terms are, the external Zeeman field 
\begin{equation}
{\bf H}_{\rm Z}={\bf H}\,,
\label{E18}
\end{equation}
that we will consider constant along the device ${\bf H(r)}  =\left( {\bf B}-{\bf M(r)} \right /\mu_0$. Note that ${\bf M(r)}\times{\bf M(r)}=0$ therefore we can use without loss of generality ${\bf H}\approx {\bf B}/\mu_0$. The exchange field
\begin{equation}
{\bf H}_{\rm ex}=\frac{2A}{\mu_0 M_{\rm s}^2} \nabla^2 {\bf M(r)}\,,
\label{E19}
\end{equation}
and the DMI effective field,
\begin{equation}
{\bf H}_{\rm DMI}=-\frac{2\mathcal{D}}{\mu_0 M_{\rm s}^2} \left(\left(\nabla \cdot {\bf M(r)} \right)\hat{z} - \nabla  M_z({\bf r}) \right)  \,,
\label{E21}
\end{equation}
where $M_z({\bf r})$ is the z component of the magnetization. The magnetization strength $M_{\rm s}=|{\bf M(r_i)}|=|{\bf M(r_j)}|$ is also constant all along the device, only its orientation varies point to point.

Finally, the interaction between the magnetization field and the spin of the conduction electrons is represented by the torque field,
\begin{equation}
{\bf H}_{\rm T}=\frac{J_{\rm sd}}{\gamma_0 M_{\rm s}} {\bf m(r)}\,.
\label{E22}
\end{equation}
In order to calculate the magnetization field of the conduction electrons ${\bf m(r)}$ a model of their movement is needed, more on this is to be found below.

Furthermore, the constants of the LL equation can be written as a function of the constants of the quantum microscopic equation $A=JS^2/a_c$ and $\mathcal{D}=2 D S^2/a_c^2$ where $S=|\langle {\bf S_i} \rangle|=|\langle {\bf S_j} \rangle|$ is the mean value of the dimensionless spin.

\subsection{Conduction electrons effective Hamiltonian}
The overall torque term can be rewritten as an independent term apart from the rest of the effective fields,
\begin{equation}
T=-\frac{1}{1+\alpha^2}\frac{J_{\rm sd}}{\gamma_0 M_{\rm s}} \left( {\bf M(r)} \times {\bf m(r)} \right)\,,
\label{E23}
\end{equation}
where ${\bf m(r)}\propto\langle {\bf \hat{s}} \rangle|_{\bf r}=Tr[\hat{\rho}({\bf r}) {\bf \hat{s}}]$ is the trace of the spin of the conduction electrons evaluated at a particular position. Naturally, we need to know which is the actual density matrix of the system at each time step. To this end we calculate the effective Hamiltonian that drives the conduction electrons tracing out the degrees of freedom related to the non-itinerant electrons,
\begin{equation}
\hat{\mathcal{H}}_{\rm eff}=\langle \hat{\mathcal{H}} \rangle|_{\rm ni}=Tr[\hat{\rho} \hat{\mathcal{H}}]_{\rm ni}=\hat{\mathcal{H}}_{\rm  K}+\hat{\mathcal{H}}^{\rm  eff}_{\rm  sd}+E_{\rm  LL}(\Psi_l)
\label{E24}
\end{equation}
where $\langle \hat{\mathcal{H}} \rangle|_{\rm ni}$ is the partial trace over the non-itinerant degrees of freedom.

The resulting effective Hamiltonian is divided in three terms. The first term, is the kinetic energy term of the conduction electrons $\hat{\mathcal{H}}_{\rm K}$ that remains unmodified with respect to its definition in eq.\,\ref{E7B} because it does not depend on pinned electron spins. Therefore, non-itinerant electronic degrees of freedom just trace out for this term. The second term, is the effective sd interaction term,
\begin{equation}
\hat{\mathcal{H}}^{\rm eff}_{\rm sd}=\langle \hat{\mathcal{H}}_{\rm sd}\rangle=\frac{J_{\rm sd} S}{M_{\rm s}}\hat{s} \cdot {\bf M(r)}\,,
\label{E25}
\end{equation}
that takes the form of an external Zeeman field for the conduction electrons where the role of the external field is taken by the ferromagnet magnetization. And the third term, is the energy provided by non-itinerant electrons terms of the Hamiltonian $E_{\rm LL}$. 
This last term depends on the states of the conduction electrons that at the same time depend on the spin configuration of the atomic ones. 

One key point of this work is the assumption that the characteristic time scale of the evolution of the conduction electron states is faster than the dynamics of the ferromagnet magnetization ${\bf M(r)}$. Therefore $E_{LL}$ will be assumed constant, decoupling both kind of degrees of freedom in an analogous way to the Born-Oppenheimmer approximation. The constant term in eq.\,\ref{E24} Hamiltonian can be ruled out shifting the origin of energies thus leading to the final form of the effective Hamiltonian,
\begin{equation}
\hat{\mathcal{H}}_{\rm eff}=\frac{{\bf \hat{p}}^2}{2m}+ {\bf \Delta}_B({\bf r}) \cdot {\bf \hat{\sigma}}\,,
\label{E25B}
\end{equation}
where ${\bf \Delta}_{\rm B}({\bf r})=\frac{J_{sd} S}{2} \frac{{\bf M(r)}}{M_{\rm s}}$. 

In this approximation, the conduction electron wavefunctions are assumed to undergo adiabatic evolution. This way, we neglect non-equilibrium effects \cite{Popescu} but this is correct provided that $|{\bf m(r)}|_{\rm max}/M_{\rm s}<<1$ where $|{\bf m(r)}|_{\rm max}$ is the maximum value in magnitude of the conduction electrons magnetization at any point.  In a physical system where the ferromagnet is in contact with, for example a wide iron slab, magnetization ratios are usually around $|{\bf m(r)}|_{\rm max}/M_{\rm s}\approx 10^{-2}$ and this value will be even smaller in narrow nanowires due to transverse confinement. 

Magnetization ratios are relevant as an adiabaticity measurement because the sd interaction between the spins of both conduction and pinned electrons is an action reaction force . Therefore, both magnetizations (of the conduction and pinned electrons) feel the same torque but with opposite sign. Naturally, the one with the smaller magnetization magnitude will change at a faster rate for the same force. 

\subsection{Resolution method.}
To calculate the time evolution of the magnetization ${\bf M}({\bf r},t)$ we will numerically integrate the LL equation (eq.\,\ref{E16}) discretized in space and time. There are different methods available to this purpose, Euler, Heun or ODE45 just to cite a few\cite{LLGnumerics,ODE45}. The calculation of the torque involves the solution of a computational costly quantum model and therefore the Euler method is preferred over Runge-Kutta methods where multiple calculations of eq.\,\ref{E16} are needed in each time step.

In the adiabatic approximation, the Hamiltonian of eq.\,\ref{E25B} will be used to calculate the conduction electron eigenstates at a given time using ${\bf M}({\bf r},t)$ as an input parameter. The magnetization of the conduction electrons ${\bf m}({\bf r},t)$ is obtained as a mean value of the ensemble of the occupied electronic eigenstates (more on this below). This magnetization ${\bf m}({\bf r},t)$ is further used to calculate the torque for the LL integration finally obtaining a new ferromagnet magnetization ${\bf M}({\bf r},t+\Delta t)$ thus closing the loop. The whole time evolution of ${\bf M}({\bf r},t)$ is then obtained iterativelly.    

\begin{figure}
  \includegraphics[width=\columnwidth]{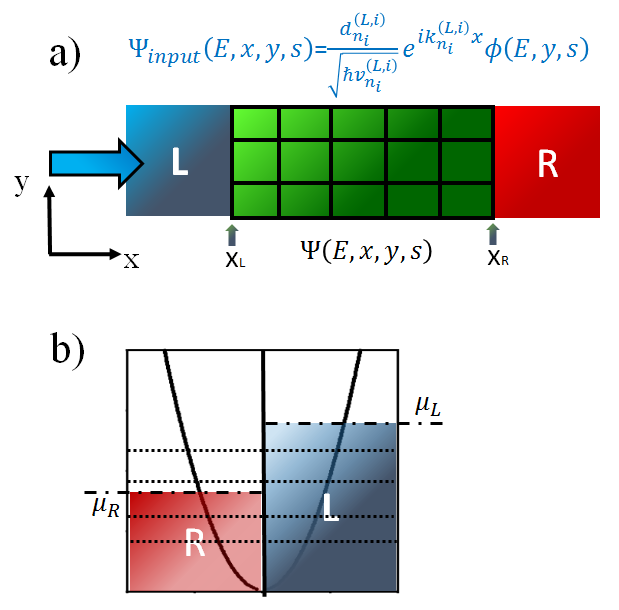}
  \caption{ a) The conduction electrons eigenstates are calculated using a quantum model on a spatial grid with the input plane waves acting as boundary conditions. b) A net charge and spin current is obtained when the chemical potential at the left and right leads are different. Note that in the single band limit the lower energy band is polarized due the magnetization of the ferromagnet.}
  \label{F1B}
\end{figure}

The wavefunctions associated with the conduction electrons are obtained as the eigenstates of  the Hamiltonian eq.\,\ref{E25B} where the interaction between conduction and pinned electrons is modeled as an external magnetic field. To calculate this eigenstates we will consider our device as a central region between two contacts (see fig.\,\ref{F1B}). This effective magnetic field will be inhomogeneous  in the central region because of the presence of a skyrmion in the ferromagnet while a constant field is assumed for the leads. The central region is discretized in the same way as the LL equation with a value of ${\bf M(r,t)}$ defined on each point of the grid. The solutions in the central region for different energies will be obtained considering eq. \ref{E25B} evaluated on each grid point using energy as a input parameter. The nanowire upper and lower boundaries in fig.\,\ref{F1B} are modeled as infinite confining potentials while the left and right boundaries are considered open contacts where the magnetic field is maintained homogeneous.

Both contacts in fig.\,\ref{F1B} are modeled as normal metals with the same effective mass as in the central region and a voltage bias may be defined between them in order to create charge and spin current. This voltage bias is introduced as difference in the chemical potentials of the left $\mu_L$ and right $\mu_R$ contacts. Incident modes from the contacts may be transmitted or reflected and therefore solutions in the contacts are linear superpositions of the asymptotic nanowire eigensolutions. The asymptotic eigensolutions are labeled by their wavenumbers because contacts are homogeneous and therefore transitionally invariant. As a consequence, the wavefunction eigensolutions at the contacts for a given energy take the form,
\begin{equation}
\label{E26}
\Psi^c(E,x,y,s) =
\sum_{\alpha,n_\alpha} 
{\frac{d^{(c,\alpha)}_{n_\alpha}}{\sqrt{\hbar v^{(c,\alpha)}_{n_\alpha}  }}
\exp{\left[ik^{(c,\alpha)}_{n_\alpha}x\right]}
\phi^{(c,\alpha)}_{n_\alpha}(y,s) 
}\; ,
\end{equation}
where $c=L,R$ labels the contact, $\alpha=i,o$ the input and output modes in each contact and $s=\uparrow,\downarrow$ the spin up and down quantum number. $d^{(c,\alpha)}_{n_\alpha}$ determines the amplitudes of the asymptotic solutions, $k^{(c,\alpha)}_{n_\alpha}$ their wavenumber and 
\begin{equation}
v^{(c,\alpha)}_{n_\alpha}=\frac{1}{\hbar}\frac{\partial E}{ \partial k^{(c,\alpha)}_{n_\alpha} }=\frac{\hbar k^{(c,\alpha)}_{n_\alpha} }{m}
\label{E26B}
\end{equation}
their group velocity . 
 
From the point of view of the conduction electrons this is a scattering problem where the skyrmion is a magnetic inhomogeneity.  To solve this problem we use an extended version of the quantum transmitting boundary method \cite{Lent} as presented in refs. \cite{Osca1,Osca2}. The overall system is described by a closed system of linear equations,
\begin{widetext}
\begin{eqnarray}
\label{E27}
\left(\hat{\mathcal{H}}_{\rm eff}-E\right) \Psi(E,x,y,s) &=& 0\; , 
\qquad\qquad\qquad\qquad\qquad\qquad\qquad\quad\quad\;\, 
 (xy)\in C \; ,	\;\\
\label{E28}
\Psi(E,x,y,s) 
-
\sum_{n_o}
{\frac{d^{(c,o)}_{n_o}}{\sqrt{\hbar v^{(c,o)}_{n_o}  }}
\exp{\left[ik^{(c,o)}_{n_o}\right]}
\phi^c_{n_o}(E,y,s)
}
&=& 
\sum_{n_i}
{\frac{d^{(c,i)}_{n_i}}{\sqrt{\hbar v^{(c,i)}_{n_i}  }}
\exp{\left[ik^{(c,i)}_{n_i}\right]}
\phi^c_{n_i}(E,y,s)
}\; , 
\,\quad\, (x,y,c)\in L/R \; ,\quad\\
\label{E29}
\sum_{s}\int{dy\,
\phi^{(c,o)}_{m_o}(E,y,s)^*
\, \Psi(E,x_c,y,s)}  
&-&
\sum_{n_o}
{\frac{d^{(c,o)}_{n_o}}{\sqrt{\hbar v^{(c,o)}_{n_o}  }}
\exp{\left[ik^{(c,o)}_{n_o}x_c\right]}
{\cal M}_{m_o n_o}^{(o c,o c)}(E)
}
=  \nonumber\\
 && \sum_{n_i}
{\frac{d^{(c,i)}_{n_i}}{\sqrt{\hbar v^{(c,i)}_{n_i}  }}
\exp{\left[ik^{(c,i)}_{n_i} x_c\right]}
{\cal M}_{m_o n_i}^{(o c,i c)}(E)
} \; ,
\; c\in L/R \; , \;
\end{eqnarray}
\end{widetext}
that can be solved numerically\cite{Harwell} where $x_c$ is the coordinate of the boundary $c=L,R$ and
\begin{equation}
{\cal M}_{m_\alpha n_\beta}^{(\alpha c,\beta c)}(E)
=
\sum_{s}
\int{dy\,
\phi_{m_\alpha}^{(\alpha,c)}(E,y,s)^*
\phi_{m_\beta}^{(\beta,c)}(E,y,s)
}\; .
\end{equation}

The first equation is just the Schr\"{o}dinger equation with $E$ as a parameter for the central 
region while the second one represents the matching between the asymptotic leads and the central region. Output modes are at the left hand side of the equation while input modes are at the right hand side. Input modes amplitudes are parameters while output modes amplitudes are unknowns to be determined. The purpose of the third set of equations \ref{E29} is to close the system of equations evaluating the strength of the overlap between the different asymptotic solutions.

The total magnetization of the conduction electrons ${\bf m(r)}$ is obtained by integrating the magnetization of each eigenstate occupied by an electron due to an active incident mode, $d_{n_{(c,i)}}=1$ (global phase is arbitrary). This is equivalent to the local trace where the Fermi-Dirac distribution takes the role of the probability for each pure state. At zero temperature we consider an incident mode active if it is below its contact Fermi energy, that is:
\begin{widetext}
\begin{equation}
\begin{split}
{\bf m}(x,y)=&-\gamma \frac{\hbar}{4\pi}\sum_{n_i}\int_0^\infty{ \left(\,f(\mu_L)\,\Psi_{n_i}^*(E,x,y)\,{\bf \hat{\sigma}}\,\Psi_{n_i}(E,x,y)\,\right)}\,dE 
 -\gamma \frac{\hbar}{4\pi}\sum_{n_i}\int_0^\infty{ \left(\,f(\mu_R)\,\Psi_{n_i}^*(E,x,y)\,{\bf \hat{\sigma}}\,\Psi_{n_i}(E,x,y)\,\right)}\,dE \\
=&-\gamma \frac{\hbar}{4\pi}\sum_{n_i}\int_0^{\mu_L}{ \left(\,\Psi_{n_i}^*(E,x,y)\,{\bf \hat{\sigma}}\,\Psi_{n_i}(E,x,y)\,\right)}\,dE 
 -\gamma \frac{\hbar}{4\pi}\sum_{n_i}\int_0^{\mu_R}{ \left(\,\Psi_{n_i}^*(E,x,y)\,{\bf \hat{\sigma}}\,\Psi_{n_i}(E,x,y)\,\right)}\,dE 
\end{split}
\end{equation}
\end{widetext}
where ${\bf \hat{\sigma}}=\left({\bf \hat{\sigma}_{x}},{\bf \hat{\sigma}_{y}},{\bf \hat{\sigma}_{z}} \right)$ and ${\bf \hat{\sigma}_{x,y,z}}$ are the corresponding Pauli matrices. Note that different magnetizations are obtained if both contacts are in equilibrium creating zero net charge and spin currents or if a potential bias is applied between them like in Fig. \ref{F1B}b. Contributions to the magnetization calculation from bound states of the electrons attached to a skyrmion are neglected. In this case, these states are not propagating therefore these electrons magnetization will be oriented in the local magnetization direction thus providing zero torque. In general, there is also the possibility of states able to create closed loops of current without any input or output from the contacts. These close loops could arise, for example, from edge states or circular motion caused by orbital effects. However, in this simple metal model such effects are not present.

\section{Results}
\label{S3}

\subsection{Skyrmion electron blockade in the single band limit} 
\begin{figure}
  \includegraphics[width=\columnwidth]{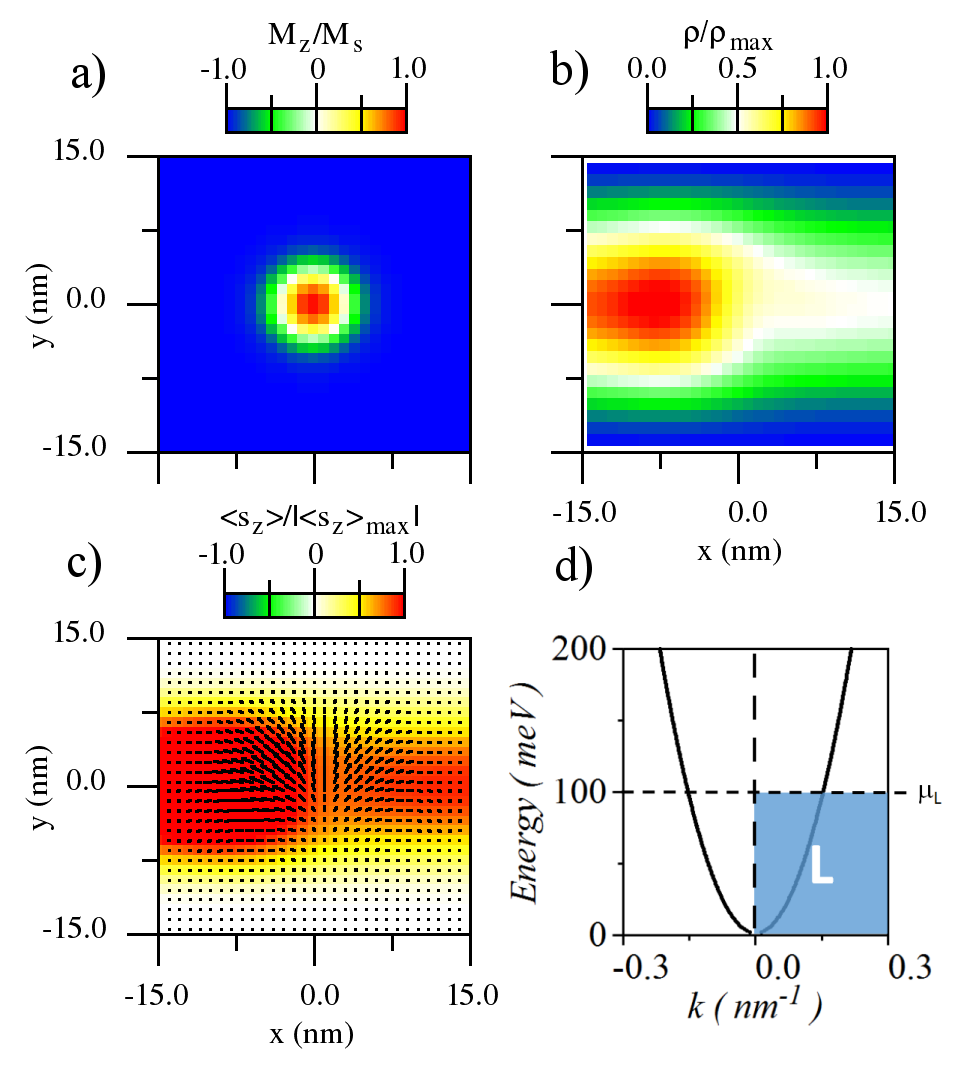}%
  \caption{a) Magnetization field of the pinned electrons. A skyrmion structure of around $10\;{\rm nm}$ diameter is observed in the center. The interface parameters used are $J=10\;{\rm meV}$, $D=1.256\;{\rm meV}$, $\hbar\gamma_0 H_z=10\;{\rm meV}$, $S=10$ and $a_c=0.1\;{\rm nm}$. Resolution of the numerical discretization $\Delta_x=\Delta_y=1\;{\rm nm}$.  b) Density of states of the conduction electrons flowing from the left lead when the skyrmion of figure a) is present. In this figure $Jsd=9.0\;{\rm meV}$, $m/m_e=0.013$ and $\mu_L=100\;{\rm meV}$ where $m_e$ is the bare electron mass. Note the rise of density at the left of the skyrmion position due the blockade effect in the flow of electrons caused by the skyrmion. c) $z$ component of the itinerant electrons magnetization (in color) and x,y components as a vector. d) Band structure and Fermi energy on the leads for the figures a),b) and c). This is the single band limit case where only incident electrons flowing from the left lead are considered. }
  \label{F2}
\end{figure}

In fig.\,\ref{F2}a the magnetization field {\bf M(r)} hosting a skyrmion is indicated. We use a spatial discretization of $1\;{\rm nm}\,\times\,1\;{\rm nm}\,\times\,1\;{\rm nm}$ for a nanowire of $L_y=25\;{\rm nm}$ wide. The ferromagnet is allowed to be larger in order to avoid border effects and it is limited by open boundary conditions. We consider a central region of length $L_x=30\;{\rm nm}$ connected to two translationally invariant infinite contacts. The interface between the ferromagnet and the nanowire is considered to be wide enough to hold an uniform electron density of 1 electron by each atomic layer $a_c=0.1\;{\rm nm}$. No confinement has been considered in the $z$ direction where electrons are assumed to decay smoothly. This resolution is fine enough with respect to the atomic length to allow the use of the LL equation with continuous fields while at the same time it is coarse enough to keep the computational cost of the quantum model reasonable. The same discretization is used for the fields ${\bf M(r,t)}$ and ${\bf m(r,t)}$ in the LL equation and in the quantum model. This resolution is coarse in comparison to simulations in literature\cite{Fert,Tomasello} but we have tested the robustness of the results comparing the torque fields at $t=0$ with their finer resolution counterparts and running higher resolution but shorter simulations of skyrmion movement. The rest of the physical parameters are discussed in fig.\,\ref{F2}a caption.
 
Physical parameters are tuned\cite{Phases} to obtain a skyrmion of radius $R\approx 5\;{\rm nm}$. The width of the nanowire is such that we are in the single band limit while the value of the skyrmion radius is constrained by the nanowire width. The conduction electron density of states 
(DOS) is shown in fig.\,\ref{F2}b in presence of the skyrmion plotted in fig.\ref{F2}a while the spin angular momentum is shown in fig.\,\ref{F2}c. The nanowire dispersion relation in the leads is displayed in fig.\,\ref{F2}d. The wire width $L_y$, the sd interaction $J_{\rm sd}$, the external magnetic field strength $H$ and the ferromagnet strength ($S=10$) have been chosen in order to obtain an ideal single band parabolic dispersion with origin at zero energy as shown in fig.\,\ref{F2}d while still having a physically plausible set of parameters. Only one band is considered, if multiple bands where shown the second band will appear at around $100\;{\rm meV}$. 

The left chemical potential $\mu_L=\,100\;{\rm meV}$ has been also chosen to match the limit of the single band regime. We can see that the electron DOS is altered by the presence of the skyrmion with an higher electron density at the left side of the skyrmion than at the right side. This also affects the spatial distribution of the magnetic moments in Fig. \ref{F2}c that is also larger because the larger electron density at the left of the skyrmion. This is happening because the partial reflection of the electron modes from the left contact caused by the magnetic inhomegenity that the skyrmion represent to the electrons. 

This electron blockade is very different to what occurs in the classical model \cite{Torque,Tomasello} where an adiabatic approximation in the sense of near detachment between the pinned and the conduction electrons magnetization is assumed. In the classical model the conduction electron magnetization ${\bf m(r)}\propto \langle {\bf \hat{s}} \rangle|_{\bf r}$ is assumed to follow in an approximated way the pinned electrons magnetization ${\bf M(r)}$ causing zero torque in first order approximation. Only second order non-adiabatic terms are responsible for generating the torque. This is very different in what is happening here in the single band limit where the electrons magnetization is altered but does not follow the skyrmion orientation while they may also be partially reflected.

\begin{figure}
  \includegraphics[width=\columnwidth]{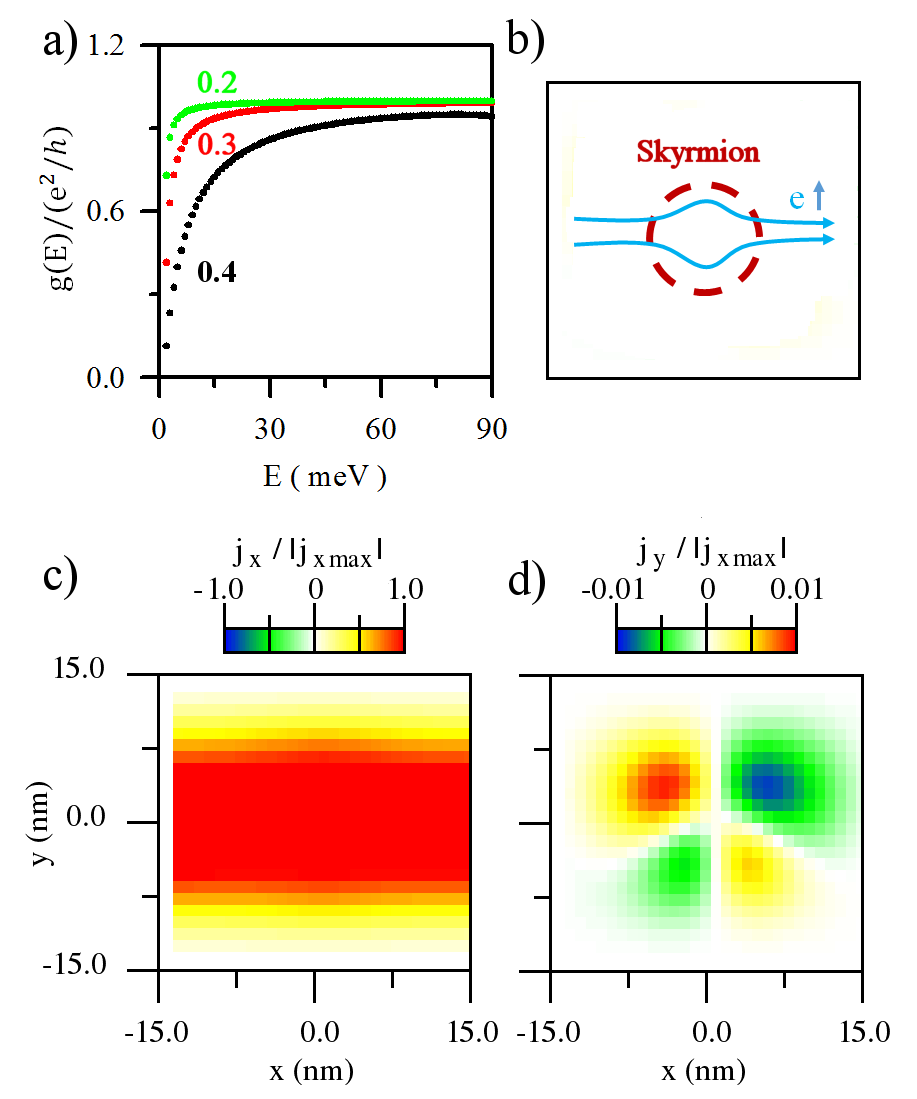}%
  \caption{a) In black, conductance of the current  for the skyrmion configuration in fig.\,\ref{F2}a.  In color, conductance for configurations with different ratios of skyrmion size in relation to the nanowire width $2R/Ly$. Size ratios are shown near the plots. Different configurations are obtained tuning $J/D$ and $H$ while maintaining $Ly$ constant. b) Schematic for the particle current in the nanowire in the single band regime in presence of a skyrmion c) $x$ component of the particle current with a larger chemical potential in the left than in the right contact, $\mu_L=100\;{\rm meV}$ and $\mu_R=0\;{\rm meV}$. The rest of the parameters are the same than in fig.\,\ref{F2}a. d) $y$ component of the particle current in the same case than c).}
  \label{F5}
\end{figure}

We obtain the conductance in the leads as
\begin{equation}
g(E)=\frac{e^2}{h}T(E)\,,
\end{equation}
where $T$ is the transmission probability at energy $E$. The electron blockade can be seen in the conductance (see fig. \ref{F5}a) where it is most
notable for lower energy values while this blockade is almost a negligible effect for larger values. The skyrmion completely blocks the lower energy states of small wavenumber with a perfect reflection of those modes while it is completely transparent for higher energy modes. Additionally, the relationship between the size of the skyrmion and the width of the nanowire also plays an important role. The increase in conductance is faster for the smaller ratios between skyrmion size and nanowire width. Therefore, this blockade can not be measured in large metallic slabs but only in very narrow wires. 

On the other hand, the local current is almost homogeneous in longitudinal $x$ direction while it has a parabolic distribution in the $y$ direction due transverse confinement. This is shown in figs.\,\ref{F5}c and\,\ref{F5}d. The $y$ component of the current exists only around the skyrmion position and it is two orders of magnitude less that the longitudinal component. In general, the particle current bends a little around the skyrmion  in an assymetrical way  therefore creating a momentum transfer between the electron current and the skyrmion. 

\subsection{Torque scaling considerations}
\begin{figure}
  \includegraphics[width=\columnwidth]{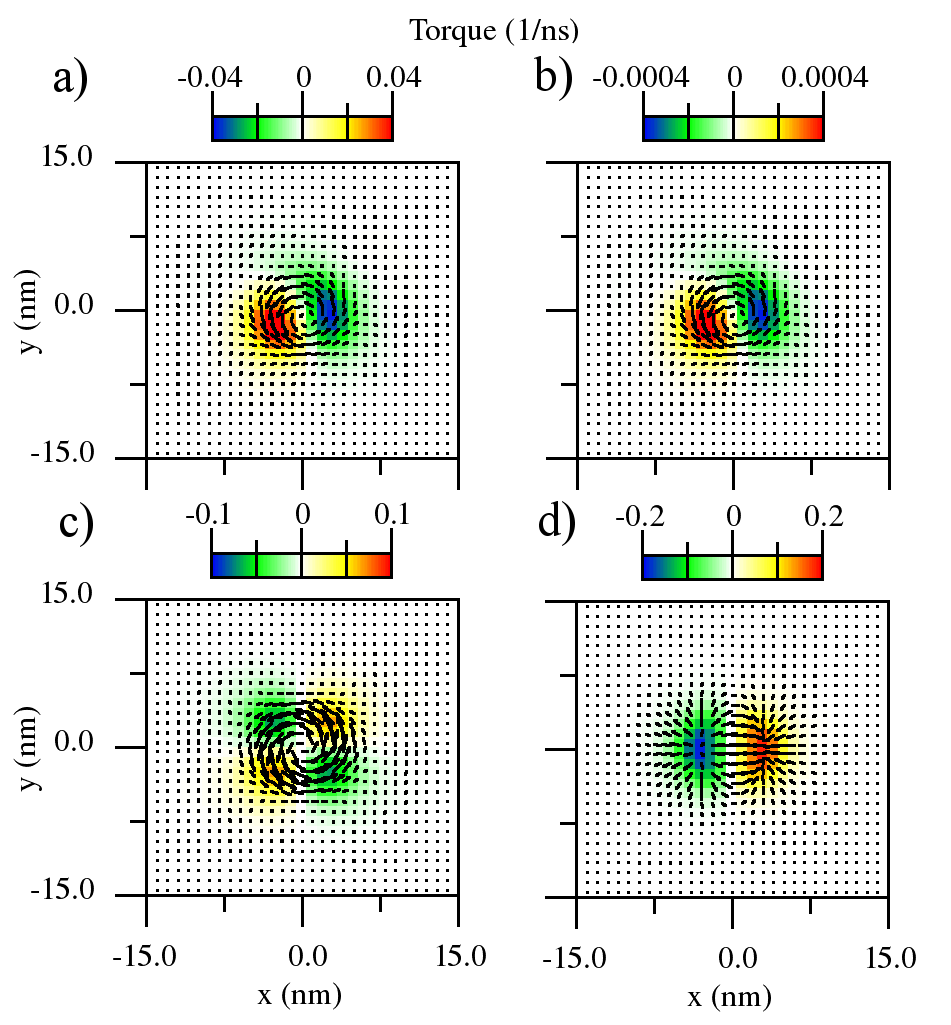}%
  \caption{Torque field as defined in eq.\,\ref{E23} for the skyrmion depicted in fig.\,\ref{F2}a. b) Same as a)  but  $J_{sd}=0.09\;{\rm meV}$, $m^*=1.3$ and $\mu_L=1\;{\rm meV}$. c) Same than in a) but for the case of zero net charge and spin currents. This is $\mu_L=\mu_R=100\;{\rm meV}$.  d)Torque provided by the classical model using the same amount of current $j_x=120\;{\rm MA/cm^2}$  that the obtained in a). Current diffusion effects have been neglected and only the ballistic torque term has been considered, that is $b_j=1$ and $c_j=0$.}
  \label{F3}
\end{figure}

Rewriting the LL equation (eq.\,\ref{E16}) and the conduction electrons effective Hamiltonian (eq.\,\ref{E25B}) as function only of dimensionless magnetizations ${\bf M(r)}/M_{\rm s}$ and ${\bf m(r)}/|m(r)|$  makes scaling relations easier to spot. The shape of the torque for a skyrmion of a given size can be maintained constant for different sd interaction strengths provided the factor $R=J_{sd} S m Ly^2$ is maintained constant. This constant is proportional to the ratio between the magnetic $J_{\rm sd} S /2$ and confinement energies $E_0=\pi^2 \hbar^2/2 m L_y^2$. As a consequence, if $J_{\rm sd}$ and $m$ are changed while maintaining $S$ constant for a given $R$ the nanowire dispersion relation is the same but for a re-scaled energy axis therefore the torque is also re-scaled in the same amount.  Note that the values of the chemical potentials in the contacts must be re-scaled too with the same amount as the energy axis. This way, the value of the Fermi wavenumber is maintained. This is shown in the comparison between figs.\,\ref{F3}a and \ref{F3}b where for two different sd interaction strengths but a common $R$ factor the same torque profile is obtained in both figures but two orders of magnitude apart. The skyrmion velocity is proportional to the torque, therefore we can infer the skyrmion velocities for a set of parameters from a single simulation. In that regard, the factor $R$ characterizes the interface determining the shape of the torque for different scales that lead to the same skyrmion movement but with different velocities.

On the other hand, if only $J_{sd}$ an $S$ are changed while maintaining the confinement energy $E_0$ constant (avoiding re-scaling of the energy bands ) then the torque strength is also maintained constant because it only depends on the product $J_{sd} S$ as shown in eq.\,\ref{E25B}. However, the skyrmion velocity will also be different for different values of $S$, the larger the $S$ the slower the dynamics of the skyrmion. Therefore a long simulation for a large $S$ but small $J_{sd}$  is equivalent to a shorter simulation of smaller $S$ where $J_{sd} S= J_{sd}' S'$. Also note that changes in $S$ will require the same proportional increase of the external magnetic field $H$ in order to keep the device in the skyrmion phase \cite{Phases}.

\subsection{Torque symmetry and skyrmion movement}
\label{subC}
A nanowire in equilibrium has no potential bias between the left and right contacts ($\mu_L=\mu_R$) therefore a zero net charge and spin current goes through the nanowire. The torque created in the quantum single band limit with zero net charge and spin current does not lead to zero torque. This is an important point of this paper because it is  different from the classical models\cite{Fert,Tomasello,Torque} where the torque becomes zero with zero net spin current. 

In figure \ref{F3}c we can see the resulting torque for the case where both contacts chemical potentials are equal $\mu_L=\mu_R=100\;{\rm meV}$. In a quantum model for transport a zero net current in a nanowire means an equal amount of right-going $k>0$ and left-going $k<0$ occupied electronic modes. In this case, a potential bias between the leads is still required to create skyrmion movement and the equilibrium case with zero net current still cancels skyrmion movement. However, differently than in the classical model the underlying reason here is not the cancellation of the torque but the symmetry of it. 

\begin{figure}
  \includegraphics[width=\columnwidth]{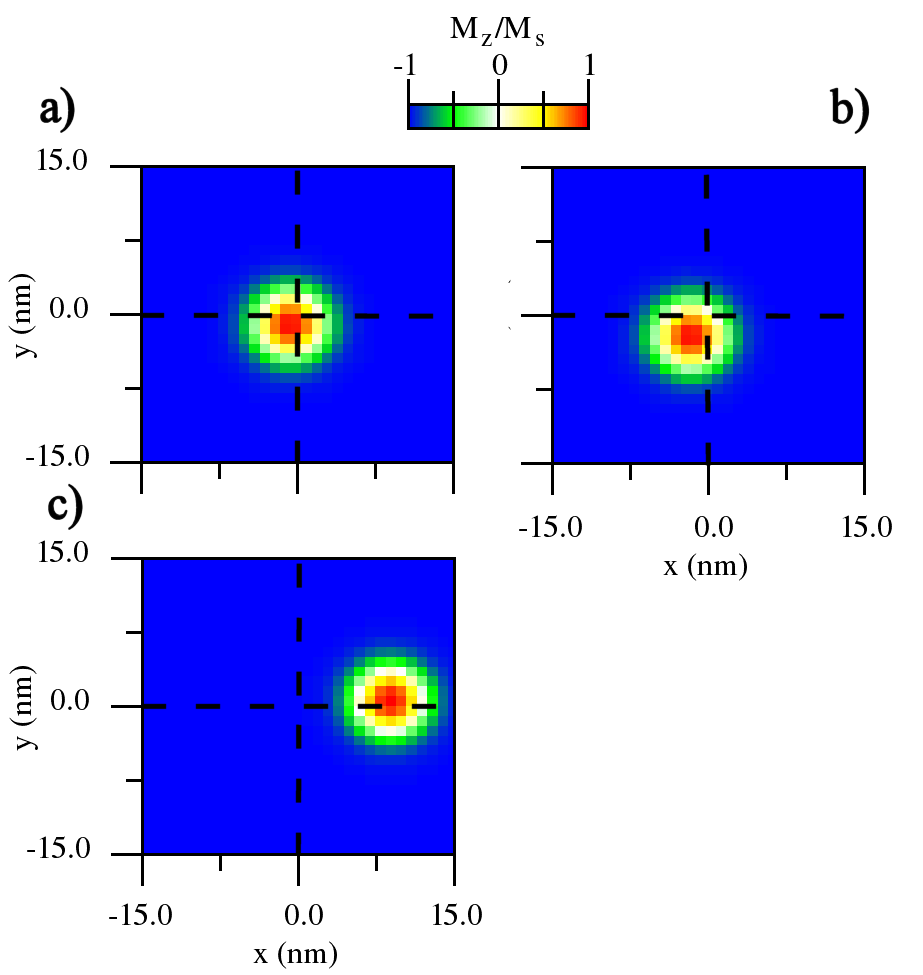}%
  \caption{a) and b) Skyrmion position after $5\;{\rm ns}$ and $10\;{\rm ns}$ of simulation respectively. The interface parameters in this case are are $J=\,10\;{\rm meV}$, $D=\,1.256\;{\rm meV}$, $\hbar\gamma_0	H_z=10\;{\rm meV}$, $S=2$, $Jsd=\,45.0\;{\rm meV}$ and $a_c=1.0\;{\rm nm}$. This results is equivalent to a $t=25\;{\rm ns}$ and $t=50\;{\rm ns}$ of simulation time for $S=10$, $Jsd=9.0\;{\rm meV}$ like in a). The resolution of the numerical discretization is $\Delta_x=\Delta_y=1\;{\rm nm}$. c) Skyrmion position after $5\;{\rm ns}$ of simulation using the classical model with the parameters and the same quantity of current density $j_x=120\;{\rm MA/cm^2}$  than in a) and b). Only the ballistic term of the torque has been considered. Therefore, $b_j=1$ and $c_j=0$.}
  \label{F6}
\end{figure}

Right-going $k>0$ electronic modes create an asymmetric torque field like in fig. \ref{F3}a. If these modes are the only ones active then skyrmion movement is created in the transverse top to bottom direction. This movement can be seen for different times in figures \ref{F6}a and \ref{F6}b. This is also different on comparison with the classical model in a STT scenario where a net right-going current will drive the skyrmion also in the right direction (see fig.\,\ref{F6}c ). The asymmetry is created by the recoil of the conduction electron spin caused by the sd interaction with the skyrmion magnetic structure. That is, the electron spin is different in the left side of the skyrmion than in the right because the skyrmion presence. Left-going modes will create the same torque but inverted around the $x=0$ axis because of the longitudinal symmetry of the device. Therefore when the same number of modes are active in both contacts because $\mu_L=\mu_R$ the resulting torque is symmetric and all the forces cancel therefore no skyrmion movement is produced.

The classical model of the torque effective field for STT used in literature is
\begin{equation}
{\bf H_T}=\frac{b_j}{\gamma_0}\frac{P\mu_B}{e g  M_{\rm s}^3}j_0  \left( {\bf M(r)} \times \nabla {\bf M(r)} \right)+\frac{c_j}{\gamma_0}\frac{P\mu_B}{e g M_{\rm s}^3}j_0 \nabla{\bf M(r)}\,,  
\label{E100}
\end{equation}
where $\mu_B$ is the Bohr magneton, $e$ the electron charge, $g\approx2$ its gyromagnetic factor and $P$ is the electronic polarization. The rest of the variables have the same meaning as above. This model of the torque contains two terms, a "ballistic" term multiplied by the constant $b_j\approx 1/1+\xi^2$ and a "diffusive term" multiplied  by $c_j\approx \xi/1+\xi^2$ . The first arises from the presence of a ballistic current in the heavy metal while the second torque term arises by the presence of diffusion effects on that current. The coefficient $\xi=\tau_{ex}/\tau_{sf}$ is calculated as the ratio between $\tau_{ex}=1/J_{sd} S$ and the spin flip relaxation time. Zero impurities in the metal imply an infinite spin flip time and therefore $\xi=0$ and $c_j=0$.

In order to compare the classical model with the single band quantum model we assume a purely ballistical nanowire $bj=1$, $c_j=0$ and also perfect polarization $P=1$. As shown in figs.\,\ref{F6}b and \ref{F6}c  skyrmion velocity is larger for the same amount of current $j_x=120\;{\rm A/nm^2}$ in the classical model than in the quantum model. Furthermore, the shape of the torque field of the quantum model in fig.\,\ref{F3}a is very different than the torque field of the classical model in fig.\,\ref{F3}d. Therefore, skyrmions move very differently in both cases, skyrmions move  in the longitudinal direction in the classical model but in the transverse direction in the quantum one. 

\begin{figure}
  \includegraphics[width=\columnwidth]{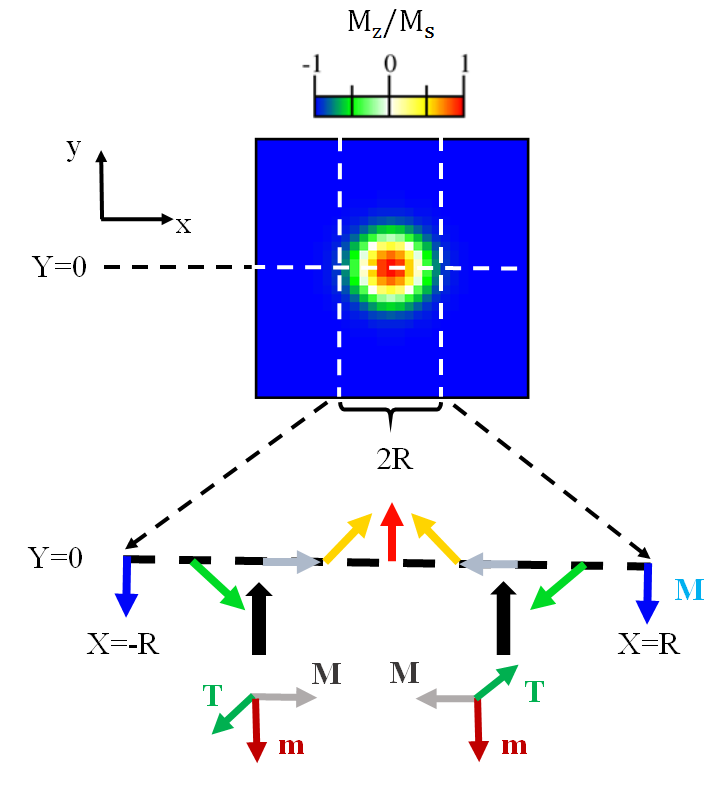}%
  \caption{Schematic of how the torque is created between a skyrmion and the itinerant electrons of a metallic very narrow nanowire in the one band limit.}
  \label{F4}
\end{figure}

We can see in fig. \ref{F4} how for the single band quantum model the magnetization ${\bf m}$ of the conduction electrons is mainly pointing upwards but the skyrmion magnetization ${\bf M}$ changes direction. Fig. \ref{F4} provides a schematic of a N\'eel skyrmion structure in a cut that goes trough its center. The cross product between conduction electrons and skyrmion magnetizations gives rise to a torque field that points mainly in the $x-y$ plane. The $z$ components of the torque arise because of the recoil between both magnetic moments, the one of the itinerant electrons and the one of the skyrmion in the ferromagnet. In the classical model the reported mechanism is different as the electron follows the skyrmion magnetic moment producing zero torque on first order approximation and torque arises from second order terms.

For the same polarization $P=1$ the single band quantum approach is less efficient in moving a skyrmion than the classical model.
The single band quantum model moves the skyrmion at a velocity around $v\approx 0.3\;{\rm m/s}$ for $S=2$ in a $10\;{\rm ns}$ simulation.  For the same amount of charge current $j_x=\,120\;{\rm A/cm^2}$ the velocity obtained from the classical model is around $v\approx 1.5\;{\rm m/s}$ using the parameters considered in fig.\ref{F6}. If $J_{sd}S$ is maintained constant these results are equivalent to a $v\approx 0.06\;{\rm m/s}$ and $v\approx 0.3\;{\rm m/s}$ in the quantum and classical models respectively in larger $50\;{\rm ns}$ simulations with $S=10$ . 

Our interpretation is that the classical model implicitly accounts for the torque interaction of many electron bands instead of only one band like in the single band quantum approach. However, wide metallic slabs with many electron bands can not be perfectly polarized to $P=1$  but $P<1$ instead. Therefore, the many electron classical limit will not benefit by its better efficiency in creating torque because it is less efficient producing spin current from charge current. Considering a realistic polarization of $P\approx0.2$ both models will give similar skyrmion velocities for the same amount of current but with different skyrmion movement directions.

\subsection{Skyrmion movement with zero current}
One of the more notable aspects of the single band limit is that it is possible to move a skyrmion even with a zero charge (and spin) current, this is, spending zero power to maintain the skyrmion in movement. As explained above, in the quantum model the same chemical potential in both contacts leads to a symmetric non-zero torque field. The imbalance between terminals creates an asymmetry in the torque that allows for the skyrmion movement but this is not the only mechanism available to create this asymmetry. 

\begin{figure}
  \includegraphics[width=\columnwidth]{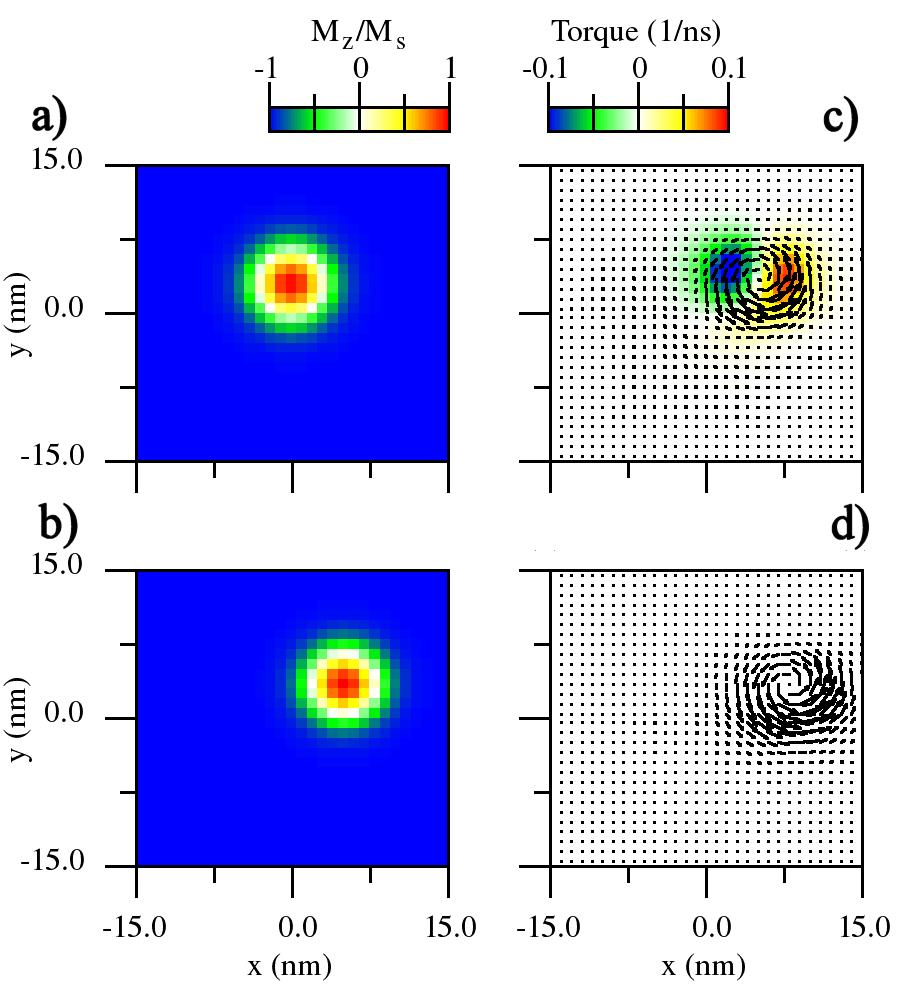}%
  \caption{a) and b) Initial skyrmion position at $y=3\;{\rm nm}$ and after $t=10\;{\rm ns}$ of simulation for the same parameters than in Fig. \ref{F6}a). c) Corresponding torque to the simulation after $t=10\;{\rm ns}$. d) Torque of a skyrmion movement simulation at $t=10\;{\rm ns}$ for the same parameters than a)
but using a synthetic unperturbed model for the conduction electrons.}
  \label{F7}
\end{figure}

As show in fig.\,\ref{F4} in the single band limit the resulting torque field is essentially the result of a cross product between the downward magnetic moment of the impinging conduction electron and the skyrmion magnetization. This way a mainly in plane torque field is created like the one shown in fig.\,\ref{F3}c where the largest torque strength is obtained at the point where the skyrmion magnetization is also pointing in plane. When both contacts are in equilibrium  the right pointing torque on the upper edge of the skyrmion and the left pointing torque on its lower edge are equal. Consequently the net torque and therefore the net movement of the skyrmion are zero as discussed in section \ref{subC}. However, if the skyrmion is placed near one of the nanowire edges ( as depicted in fig. \ref{F7}a ) the upper and lower torques become imbalanced because the smaller electronic DOS near the edge of the nanowire. If the torque at the lower edge of the skyrmion pointing to the right direction is larger
than the one on the upper edge then a net skyrmion movement in that direction arises as shown in figs.\,\ref{F7}b and \ref{F7}c. 
 
Note that the source of the asymmetry in the torque field is due only to DOS variation. Therefore, this movement can be explained without considering the effect of the torque on the conduction electrons (not depicted in fig. \ref{F4}). Conduction electron spin creates a torque on the skyrmion but they are also affected by the same torque with opposite sign. In fig.\,\ref{F7}d we use a simpler model of unperturbed electrons obtaining a similar skyrmion movement than in figs.\,\ref{F7}b and \ref{F7}c.  This new model uses the magnetic moment field of the itinerant electrons neglecting the effect of the skyrmion on them but not the other way around.

We can see comparing the torque from the quantum model in fig.\,\ref{F7}c with the one of the new model in fig.\,\ref{F7}d that a similar field and skyrmion dynamics are obtained although there is no non-zero $z$ components of the torque in the new model because electrons do not change their spin orientation across the wire. In the former case the electron recoil diminishes a bit the skyrmion velocity and it is a source of losses.

\subsection{Skyrmion torque with many conducting bands, classical limit}
\begin{figure}
  \includegraphics[width=\columnwidth]{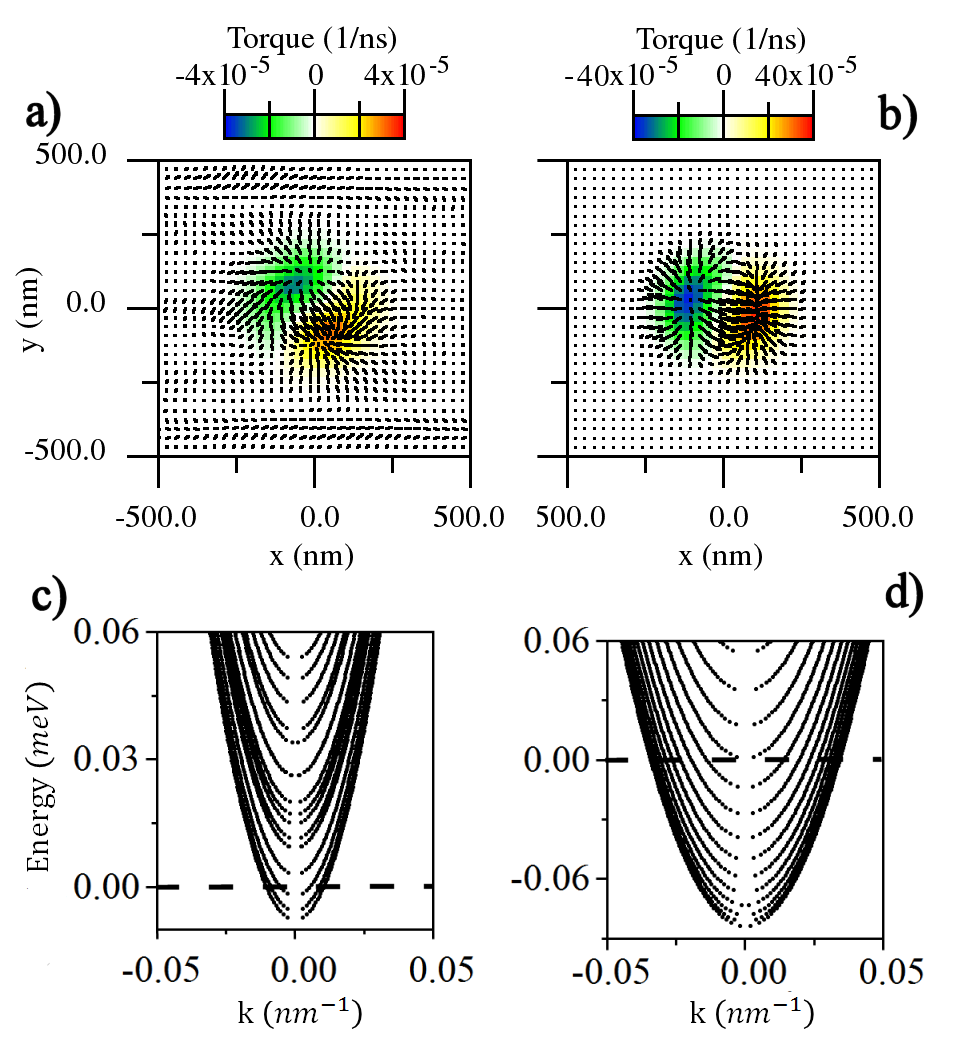}%
  \caption{a) Ensemble torque calculated with the quantum model for $n=15$ active bands in a $L_y=1\;{\rm \mu m}$ width nanowire with contacts $\mu_L=\,0.06\;{\rm meV}$ and $\mu_R=-0.01\;{\rm meV}$. The  rest of the interface parameters are $J=13.63\;{\rm meV}$, $D=0.086\;{\rm meV}$, $\hbar \gamma_0 H_z=0.7\;{\rm neV}$, $S=1.7$, $Jsd=0.01\;{\rm meV}$, $m/m_e=0.53$ and $a_c=0.316\;{\rm nm}$ where $m_e$ is the bare electron mass. The resolution of the numerical discretization is $\Delta_x=\Delta_y=15\;{\rm nm}$.  b) Ensemble torque calculated with the quantum model for $n=13$ with the same interface parameters than a) but $J_{sd}=0.2$ and $\mu_R=-0.09\;{\rm meV}$ . c) Dispersion relation for a). d) Dispersion relation for b) }
  \label{F8}
\end{figure}

\begin{figure}
  \includegraphics[width=\columnwidth]{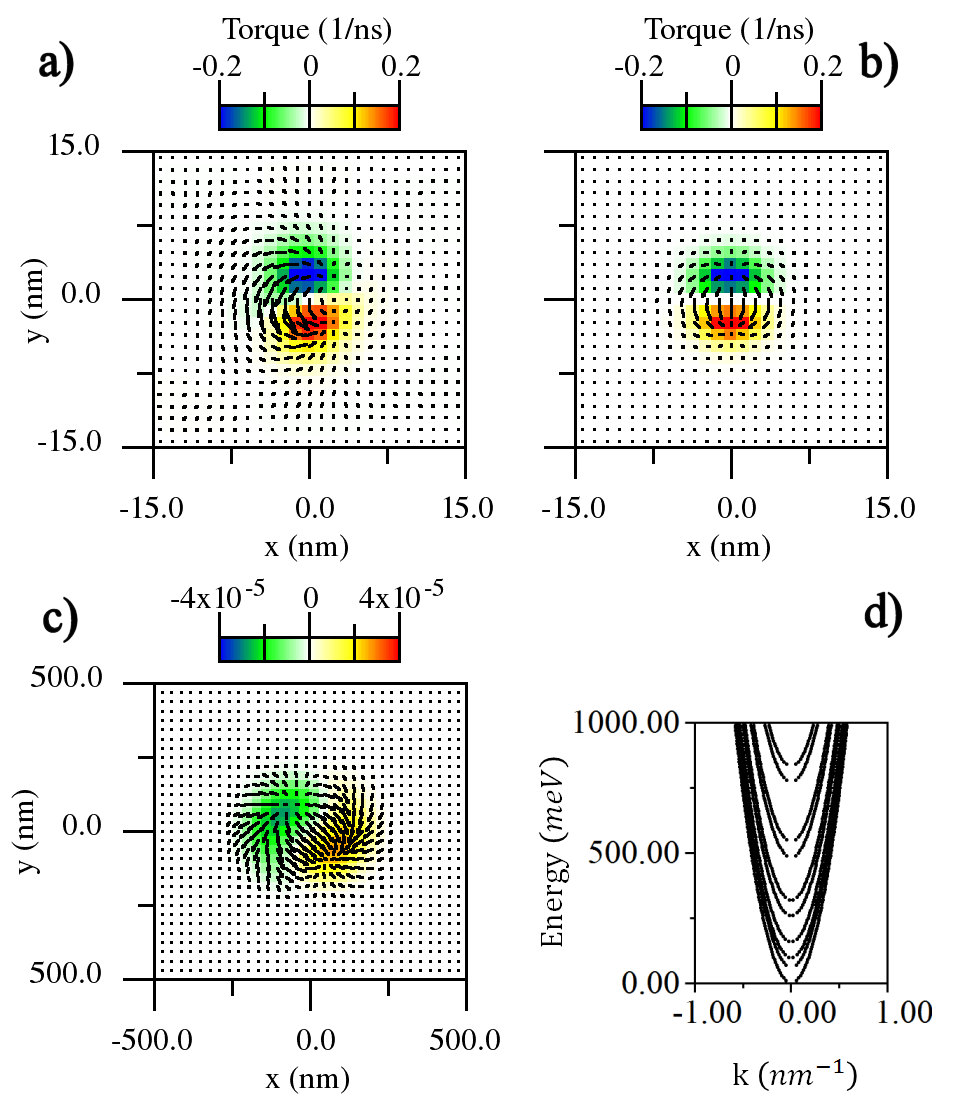}%
  \caption{a) Ensemble torque calculated with the quantum model for $n=10$ active bands in a $L_y=30\;{\rm nm}$ width nanowire with contacts $\mu_L=1\;{\rm eV}$ and $\mu_R=0\;{\rm meV}$. The  rest of the interface parameters in this case are are $J=10\;{\rm meV}$, $D=1.256\;{\rm meV}$, $\hbar \gamma_0 H_z=0.2\;{\rm meV}$, $S=2.0$, $Jsd=0.01\;{\rm meV}$, $m/m_e=0.013$ and $a_c=0.1\;{\rm nm}$ where $m_e$ is the bare electron mass. The resolution of the numerical discretization is $\Delta_x=\Delta_y=0.6\;{\rm nm}$. b) Torque calculated with the classical model for the same interface parameters than in a) where only current diffusion is present, that is $b_j=0$ and $c_j=1.0$. Current parameter $j_x=40\;{\rm MA/cm^2}$ in eq.\,\ref{E100} has been selected to match torque strengths with the quantum model. c) Torque calculated with the classical model for the same interface parameters than in fig.\,\ref{F8}a where ballistic and diffusive terms have the same weight $b_j=0.5$ and $c_j=0.5$. Current $j_x= 6\;{\rm kA/cm^2}$ has been also selected to match torque strengths. d) Dispersion relation for a).}
  \label{F9}
\end{figure}

The torque field of the classical model is recovered when many conduction bands are considered. This is the typical situation in wide slabs where the energy gap between different electronic modes is very small compared with $\mu$. This is shown in fig. \ref{F8}a where the torque field obtained from a quantum model for a few bands already takes the shape of a torque field (like the one in fig.\,\ref{F9}c) calculated with the classical model. 

Surprisingly, the torque field in fig.\,\ref{F9}c using the classical model is calculated considering that there is some amount of electron diffusion while the ensemble torque for multiple electron bands from the quantum model is calculated in a purely ballistic nanowire.  Our interpretation is that when the ratio $J_{\rm sd} S /E_0$ is not too large the skyrmion magnetization creates a small perturbation in the spin of the conduction electrons similar to the perturbation that may be created by magnetic impurities. In this regard, there is electronic diffusion because the skyrmion itself is behaving as a collection of magnetic impurities. We already showed in fig.\,\ref{F5} how the momenta of the electrons is altered around the skyrmion. 

On the other hand, with increased $J_{\rm sd}$ as in Fig. \ref{F8}b the skyrmion magnetization strength is more than a perturbation. In this regime, the conduction electrons magnetization ${\bf m(r)}$ mimics more closely the skyrmion magnetization fulfilling better the assumptions of the classical model where only the ballistic term is present (like in fig. \ref{F3}d) provided there are no impurities in the metal. 

For the same reason, this effect is not seen in wide slabs where $J_{\rm sd} S /E_0$ is large while it becomes a more important effect if we make the nanowire narrower. The torque resulting from the latter case (see fig.\,\ref{F9}a) takes a shape analogous to the torque calculated from a classical model (like in fig.\,\ref{F9}b) where the only torque present in the nanowire is caused by the diffusion term. 

The classical model is derived under the assumption that the conduction electrons spin adapt almost instantly to the skyrmion magnetization field. Intuitively, one may think that this assumption may translate to the quantum model in the form of a certain short wavelength regime. This is, larger values of $J_{\rm sd} S=1/\tau_{ex}$ may imply larger values of $k$ for a wider range of $E$. Therefore the larger $k$ the smaller the length scale an electron needs to adapt its spin orientation to the skyrmion magnetization orientation. However, this idea is wrong and the concept of electron spin instant reaction can not be carried straightforwardly between models. In figs. \ref{F8}c, \ref{F8}d and \ref{F9}d it is shown that the different dispersion relations of the conduction electrons lead to the different torque fields in  \ref{F8}a, \ref{F8}b and \ref{F9}a. In the range of energies considered the maximum wavenumber does not change that much. As a consequence, the classical concept that electrons spins relax almost instantly to an equilibrium value following roughly the skyrmion magnetization is an average statistical effect that can not be applied individually to single electrons. It is for this reason that the classical model breaks down for few conduction bands.

\section{Conclusions}
\label{S4}
The torque field created by the interaction of a skyrmion magnetization with the spin current in the single band limit is very different from the one reported in the classical model for STT. The main reason is that the classical model considers implicitly an ensemble of many electrons bands whose properties can not be translated to the behavior of individual electrons. 

The classical model limit can be recovered if many conduction bands are considered but dispersion effects may arise for a certain range of nanowire widths because of the relative strength of the sd interaction with respect the nanowire confinement energy. The quantum single band limit and the classical limit are similarly efficient in producing torque for the same quantity of charge current. The one band limit is in itself a fully polarized state and therefore it is a more efficient regime in creating spin current from a given charge current than the classical limit with polarization $P<1$. However, the classical limit is more efficient creating skyrmion movement from a given spin current that the single band limit because it is implicitly considering the effect of many electrons. 

However, each one of the limits have their advantages and limitations. On one side, there is no theoretical limitation in the amount of charge current that a metal slab can carry ( except of course the thermal resilience of the material) while there is a limitation in the maximum chemical potential that can be used while in the single band limit before further conduction bands are activated therefore breaking this limit. 

On the other hand, the single band regime offers new possibilities. Skyrmions can be detected with conductance measurements in metallic nanowires and it is possible to maintain skyrmion  movement even in zero power conditions. There are limitations in the skyrmion velocities attained by this method because the narrow range of energies where the single band limit holds before new bands are activated. However, it may be interesting for applications with strong power restrictions.

\begin{acknowledgments}
The authors thanks Lloren\c{c} Serra for useful discussion on the conduction electron quantum model. We also want to show gratitude to Dimitrios Andrikopoulos for sharing his knowledge about the available bibliography and to F.J.P. van Duijn for his comments on earlier versions of this manuscript. We acknowledge the Horizon 2020 project SKYTOP “Skyrmion-Topological Insulator and Weyl Semimetal Technology” (FETPROACT-2018-01, n. 824123). Finally, Javier Osca also acknowledges the postdoctoral fellowship provided by KU Leuven.
\end{acknowledgments} 

\bibliographystyle{apsrev4-1}
\bibliography{skybib}

\begin{thebibliography}{35}%
\makeatletter
\providecommand \@ifxundefined [1]{%
 \@ifx{#1\undefined}
}%
\providecommand \@ifnum [1]{%
 \ifnum #1\expandafter \@firstoftwo
 \else \expandafter \@secondoftwo
 \fi
}%
\providecommand \@ifx [1]{%
 \ifx #1\expandafter \@firstoftwo
 \else \expandafter \@secondoftwo
 \fi
}%
\providecommand \natexlab [1]{#1}%
\providecommand \enquote  [1]{``#1''}%
\providecommand \bibnamefont  [1]{#1}%
\providecommand \bibfnamefont [1]{#1}%
\providecommand \citenamefont [1]{#1}%
\providecommand \href@noop [0]{\@secondoftwo}%
\providecommand \href [0]{\begingroup \@sanitize@url \@href}%
\providecommand \@href[1]{\@@startlink{#1}\@@href}%
\providecommand \@@href[1]{\endgroup#1\@@endlink}%
\providecommand \@sanitize@url [0]{\catcode `\\12\catcode `\$12\catcode
  `\&12\catcode `\#12\catcode `\^12\catcode `\_12\catcode `\%12\relax}%
\providecommand \@@startlink[1]{}%
\providecommand \@@endlink[0]{}%
\providecommand \url  [0]{\begingroup\@sanitize@url \@url }%
\providecommand \@url [1]{\endgroup\@href {#1}{\urlprefix }}%
\providecommand \urlprefix  [0]{URL }%
\providecommand \Eprint [0]{\href }%
\providecommand \doibase [0]{http://dx.doi.org/}%
\providecommand \selectlanguage [0]{\@gobble}%
\providecommand \bibinfo  [0]{\@secondoftwo}%
\providecommand \bibfield  [0]{\@secondoftwo}%
\providecommand \translation [1]{[#1]}%
\providecommand \BibitemOpen [0]{}%
\providecommand \bibitemStop [0]{}%
\providecommand \bibitemNoStop [0]{.\EOS\space}%
\providecommand \EOS [0]{\spacefactor3000\relax}%
\providecommand \BibitemShut  [1]{\csname bibitem#1\endcsname}%
\let\auto@bib@innerbib\@empty
\bibitem [{\citenamefont {Skyrme}(1962)}]{Skyrme}%
  \BibitemOpen
  \bibfield  {author} {\bibinfo {author} {\bibfnamefont {T.~H.~R.}\
  \bibnamefont {Skyrme}},\ }\href@noop {} {\bibfield  {journal} {\bibinfo
  {journal} {Nucl. Phys.}\ }\textbf {\bibinfo {volume} {31}},\ \bibinfo {pages}
  {556} (\bibinfo {year} {1962})}\BibitemShut {NoStop}%
\bibitem [{\citenamefont {Donati}\ and\ \citenamefont {et. al.}(2016)}]{Topo}%
  \BibitemOpen
  \bibfield  {author} {\bibinfo {author} {\bibfnamefont {S.}~\bibnamefont
  {Donati}}\ and\ \bibinfo {author} {\bibnamefont {et. al.}},\ }\href {\doibase
  10.1073/pnas.1610123114} {\bibfield  {journal} {\bibinfo  {journal}
  {Proceedings of the National Academy of Sciences}\ }\textbf {\bibinfo
  {volume} {113}},\ \bibinfo {pages} {14926} (\bibinfo {year}
  {2016})}\BibitemShut {NoStop}%
\bibitem [{\citenamefont {Fert}\ \emph {et~al.}(2013)\citenamefont {Fert},
  \citenamefont {Cros},\ and\ \citenamefont {Sampaio}}]{Fert}%
  \BibitemOpen
  \bibfield  {author} {\bibinfo {author} {\bibfnamefont {A.}~\bibnamefont
  {Fert}}, \bibinfo {author} {\bibfnamefont {V.}~\bibnamefont {Cros}}, \ and\
  \bibinfo {author} {\bibfnamefont {J.}~\bibnamefont {Sampaio}},\ }\href
  {\doibase 10.1038/nnano.2013.29} {\bibfield  {journal} {\bibinfo  {journal}
  {Nature Nanotech}\ }\textbf {\bibinfo {volume} {8}},\ \bibinfo {pages} {152}
  (\bibinfo {year} {2013})}\BibitemShut {NoStop}%
\bibitem [{\citenamefont {Tomasello}\ \emph {et~al.}(2015)\citenamefont
  {Tomasello}, \citenamefont {Martinez}, \citenamefont {Zivieri},\ and\
  \citenamefont {et~al.}}]{Tomasello}%
  \BibitemOpen
  \bibfield  {author} {\bibinfo {author} {\bibfnamefont {R.}~\bibnamefont
  {Tomasello}}, \bibinfo {author} {\bibfnamefont {E.}~\bibnamefont {Martinez}},
  \bibinfo {author} {\bibfnamefont {R.}~\bibnamefont {Zivieri}}, \ and\
  \bibinfo {author} {\bibnamefont {et~al.}},\ }\href {\doibase
  10.1038/srep06784} {\bibfield  {journal} {\bibinfo  {journal} {Sci Rep}\
  }\textbf {\bibinfo {volume} {4}},\ \bibinfo {pages} {6784} (\bibinfo {year}
  {2015})}\BibitemShut {NoStop}%
\bibitem [{\citenamefont {Dzyaloshinsky}(1958)}]{Dzyal}%
  \BibitemOpen
  \bibfield  {author} {\bibinfo {author} {\bibfnamefont {I.}~\bibnamefont
  {Dzyaloshinsky}},\ }\href@noop {} {\bibfield  {journal} {\bibinfo  {journal}
  {J. Phys. Chem. Solids.}\ }\textbf {\bibinfo {volume} {4}},\ \bibinfo {pages}
  {241} (\bibinfo {year} {1958})}\BibitemShut {NoStop}%
\bibitem [{\citenamefont {Moriya}(1960)}]{Moriya}%
  \BibitemOpen
  \bibfield  {author} {\bibinfo {author} {\bibfnamefont {T.}~\bibnamefont
  {Moriya}},\ }\href@noop {} {\bibfield  {journal} {\bibinfo  {journal}
  {Physical review letters}\ }\textbf {\bibinfo {volume} {4}} (\bibinfo {year}
  {1960})}\BibitemShut {NoStop}%
\bibitem [{\citenamefont {Bode}\ and\ \citenamefont {et~al.}(2007)}]{Bloch1}%
  \BibitemOpen
  \bibfield  {author} {\bibinfo {author} {\bibfnamefont {M.}~\bibnamefont
  {Bode}}\ and\ \bibinfo {author} {\bibnamefont {et~al.}},\ }\href@noop {}
  {\bibfield  {journal} {\bibinfo  {journal} {Nature}\ }\textbf {\bibinfo
  {volume} {447}},\ \bibinfo {pages} {190} (\bibinfo {year}
  {2007})}\BibitemShut {NoStop}%
\bibitem [{\citenamefont {M\"uhlbauer}\ and\ \citenamefont
  {et~al.}(2009)}]{Bloch2}%
  \BibitemOpen
  \bibfield  {author} {\bibinfo {author} {\bibfnamefont {S.}~\bibnamefont
  {M\"uhlbauer}}\ and\ \bibinfo {author} {\bibnamefont {et~al.}},\ }\href@noop
  {} {\bibfield  {journal} {\bibinfo  {journal} {Science}\ }\textbf {\bibinfo
  {volume} {323}},\ \bibinfo {pages} {915} (\bibinfo {year}
  {2009})}\BibitemShut {NoStop}%
\bibitem [{\citenamefont {Huang}\ and\ \citenamefont {Chien}(2012)}]{Bloch3}%
  \BibitemOpen
  \bibfield  {author} {\bibinfo {author} {\bibfnamefont {S.~X.}\ \bibnamefont
  {Huang}}\ and\ \bibinfo {author} {\bibfnamefont {C.~L.}\ \bibnamefont
  {Chien}},\ }\href@noop {} {\bibfield  {journal} {\bibinfo  {journal} {Phys.
  Rev. Lett.}\ }\textbf {\bibinfo {volume} {108}},\ \bibinfo {pages} {267201}
  (\bibinfo {year} {2012})}\BibitemShut {NoStop}%
\bibitem [{\citenamefont {Sampaio}\ \emph {et~al.}(2013)\citenamefont
  {Sampaio}, \citenamefont {Cros}, \citenamefont {Rohart}, \citenamefont
  {Thiaville},\ and\ \citenamefont {Fert}}]{Neel1}%
  \BibitemOpen
  \bibfield  {author} {\bibinfo {author} {\bibfnamefont {J.}~\bibnamefont
  {Sampaio}}, \bibinfo {author} {\bibfnamefont {V.}~\bibnamefont {Cros}},
  \bibinfo {author} {\bibfnamefont {S.}~\bibnamefont {Rohart}}, \bibinfo
  {author} {\bibfnamefont {A.}~\bibnamefont {Thiaville}}, \ and\ \bibinfo
  {author} {\bibfnamefont {A.}~\bibnamefont {Fert}},\ }\href@noop {} {\bibfield
   {journal} {\bibinfo  {journal} {Nat. Nanotech.}\ }\textbf {\bibinfo {volume}
  {8}},\ \bibinfo {pages} {839} (\bibinfo {year} {2013})}\BibitemShut {NoStop}%
\bibitem [{\citenamefont {Rohart}\ and\ \citenamefont
  {Thiaville}(2013)}]{Neel2}%
  \BibitemOpen
  \bibfield  {author} {\bibinfo {author} {\bibfnamefont {S.}~\bibnamefont
  {Rohart}}\ and\ \bibinfo {author} {\bibfnamefont {A.}~\bibnamefont
  {Thiaville}},\ }\href@noop {} {\bibfield  {journal} {\bibinfo  {journal}
  {Phys. Rev. B}\ }\textbf {\bibinfo {volume} {88}},\ \bibinfo {pages} {184422}
  (\bibinfo {year} {2013})}\BibitemShut {NoStop}%
\bibitem [{\citenamefont {Ferriani}\ and\ \citenamefont
  {et~al.}(2008)}]{Neel3}%
  \BibitemOpen
  \bibfield  {author} {\bibinfo {author} {\bibfnamefont {P.}~\bibnamefont
  {Ferriani}}\ and\ \bibinfo {author} {\bibnamefont {et~al.}},\ }\href@noop {}
  {\bibfield  {journal} {\bibinfo  {journal} {Phys. Rev. Lett.}\ }\textbf
  {\bibinfo {volume} {101}},\ \bibinfo {pages} {027201} (\bibinfo {year}
  {2008})}\BibitemShut {NoStop}%
\bibitem [{\citenamefont {Ralph}\ and\ \citenamefont
  {Stiles}(2008)}]{RevTorques}%
  \BibitemOpen
  \bibfield  {author} {\bibinfo {author} {\bibfnamefont {D.}~\bibnamefont
  {Ralph}}\ and\ \bibinfo {author} {\bibfnamefont {M.}~\bibnamefont {Stiles}},\
  }\href {\doibase 10.1016/j.jmmm.2007.12.019} {\bibfield  {journal} {\bibinfo
  {journal} {Journal of Magnetism and Magnetic Materials}\ }\textbf {\bibinfo
  {volume} {320}},\ \bibinfo {pages} {1190} (\bibinfo {year}
  {2008})}\BibitemShut {NoStop}%
\bibitem [{\citenamefont {Zhang}\ and\ \citenamefont {Li}(2004)}]{Torque}%
  \BibitemOpen
  \bibfield  {author} {\bibinfo {author} {\bibfnamefont {S.}~\bibnamefont
  {Zhang}}\ and\ \bibinfo {author} {\bibfnamefont {Z.}~\bibnamefont {Li}},\
  }\href {\doibase 10.1103/PhysRevLett.93.127204} {\bibfield  {journal}
  {\bibinfo  {journal} {Phys. Rev. Lett.}\ }\textbf {\bibinfo {volume} {93}},\
  \bibinfo {pages} {127204} (\bibinfo {year} {2004})}\BibitemShut {NoStop}%
\bibitem [{\citenamefont {Zhang}\ \emph
  {et~al.}(2015{\natexlab{a}})\citenamefont {Zhang}, \citenamefont {Baker},
  \citenamefont {Komineas},\ and\ \citenamefont {et~al.}}]{Processor1}%
  \BibitemOpen
  \bibfield  {author} {\bibinfo {author} {\bibfnamefont {S.}~\bibnamefont
  {Zhang}}, \bibinfo {author} {\bibfnamefont {A.}~\bibnamefont {Baker}},
  \bibinfo {author} {\bibfnamefont {S.}~\bibnamefont {Komineas}}, \ and\
  \bibinfo {author} {\bibnamefont {et~al.}},\ }\href {\doibase
  10.1038/srep15773} {\bibfield  {journal} {\bibinfo  {journal} {Sci Rep}\
  }\textbf {\bibinfo {volume} {5}},\ \bibinfo {pages} {15773} (\bibinfo {year}
  {2015}{\natexlab{a}})}\BibitemShut {NoStop}%
\bibitem [{\citenamefont {Zhang}\ \emph
  {et~al.}(2015{\natexlab{b}})\citenamefont {Zhang}, \citenamefont {Ezawa},\
  and\ \citenamefont {Zhou}}]{Processor2}%
  \BibitemOpen
  \bibfield  {author} {\bibinfo {author} {\bibfnamefont {X.}~\bibnamefont
  {Zhang}}, \bibinfo {author} {\bibfnamefont {M.}~\bibnamefont {Ezawa}}, \ and\
  \bibinfo {author} {\bibfnamefont {Y.}~\bibnamefont {Zhou}},\ }\href {\doibase
  10.1038/srep09400} {\bibfield  {journal} {\bibinfo  {journal} {Sci Rep}\
  }\textbf {\bibinfo {volume} {5}},\ \bibinfo {pages} {9400} (\bibinfo {year}
  {2015}{\natexlab{b}})}\BibitemShut {NoStop}%
\bibitem [{\citenamefont {Kang}\ \emph {et~al.}(2016)\citenamefont {Kang},
  \citenamefont {Huang},\ and\ \citenamefont {Zheng}}]{Kang}%
  \BibitemOpen
  \bibfield  {author} {\bibinfo {author} {\bibfnamefont {W.}~\bibnamefont
  {Kang}}, \bibinfo {author} {\bibfnamefont {Y.}~\bibnamefont {Huang}}, \ and\
  \bibinfo {author} {\bibfnamefont {C.~e.~a.}\ \bibnamefont {Zheng}},\ }\href
  {\doibase 10.1038/srep23164} {\bibfield  {journal} {\bibinfo  {journal} {Sci
  Rep}\ }\textbf {\bibinfo {volume} {6}} (\bibinfo {year} {2016}),\
  10.1038/srep23164}\BibitemShut {NoStop}%
\bibitem [{\citenamefont {Parkin}\ \emph {et~al.}(2008)\citenamefont {Parkin},
  \citenamefont {Hayashi},\ and\ \citenamefont {Thomas}}]{IBM}%
  \BibitemOpen
  \bibfield  {author} {\bibinfo {author} {\bibfnamefont {S.~S.~P.}\
  \bibnamefont {Parkin}}, \bibinfo {author} {\bibfnamefont {M.}~\bibnamefont
  {Hayashi}}, \ and\ \bibinfo {author} {\bibfnamefont {L.}~\bibnamefont
  {Thomas}},\ }\href@noop {} {\bibfield  {journal} {\bibinfo  {journal}
  {Science}\ }\textbf {\bibinfo {volume} {320}},\ \bibinfo {pages} {190}
  (\bibinfo {year} {2008})}\BibitemShut {NoStop}%
\bibitem [{\citenamefont {Koshibae}\ and\ \citenamefont
  {Nagaosa}(2017)}]{STT1}%
  \BibitemOpen
  \bibfield  {author} {\bibinfo {author} {\bibfnamefont {W.}~\bibnamefont
  {Koshibae}}\ and\ \bibinfo {author} {\bibfnamefont {N.}~\bibnamefont
  {Nagaosa}},\ }\href@noop {} {\bibfield  {journal} {\bibinfo  {journal}
  {Scientific Reports}\ }\textbf {\bibinfo {volume} {7}},\ \bibinfo {pages}
  {42645} (\bibinfo {year} {2017})}\BibitemShut {NoStop}%
\bibitem [{\citenamefont {Iwasaki}\ \emph {et~al.}(2013)\citenamefont
  {Iwasaki}, \citenamefont {Mochizuki},\ and\ \citenamefont {Nagaosa}}]{STT2}%
  \BibitemOpen
  \bibfield  {author} {\bibinfo {author} {\bibfnamefont {J.}~\bibnamefont
  {Iwasaki}}, \bibinfo {author} {\bibfnamefont {M.}~\bibnamefont {Mochizuki}},
  \ and\ \bibinfo {author} {\bibfnamefont {N.}~\bibnamefont {Nagaosa}},\
  }\href@noop {} {\bibfield  {journal} {\bibinfo  {journal} {Nature
  Communications}\ }\textbf {\bibinfo {volume} {4}},\ \bibinfo {pages} {1463}
  (\bibinfo {year} {2013})}\BibitemShut {NoStop}%
\bibitem [{\citenamefont {Fook}\ \emph {et~al.}(2016)\citenamefont {Fook},
  \citenamefont {Gan},\ and\ \citenamefont {Lew}}]{STT3}%
  \BibitemOpen
  \bibfield  {author} {\bibinfo {author} {\bibfnamefont {H.~T.}\ \bibnamefont
  {Fook}}, \bibinfo {author} {\bibfnamefont {W.~L.}\ \bibnamefont {Gan}}, \
  and\ \bibinfo {author} {\bibfnamefont {W.~S.}\ \bibnamefont {Lew}},\
  }\href@noop {} {\bibfield  {journal} {\bibinfo  {journal} {Scientific
  Reports}\ }\textbf {\bibinfo {volume} {6}},\ \bibinfo {pages} {21099}
  (\bibinfo {year} {2016})}\BibitemShut {NoStop}%
\bibitem [{\citenamefont {Kurebayashi}\ and\ \citenamefont
  {Nagaosa}(2019)}]{QTopological}%
  \BibitemOpen
  \bibfield  {author} {\bibinfo {author} {\bibfnamefont {D.}~\bibnamefont
  {Kurebayashi}}\ and\ \bibinfo {author} {\bibfnamefont {N.}~\bibnamefont
  {Nagaosa}},\ }\href {\doibase 10.1103/PhysRevB.100.134407} {\bibfield
  {journal} {\bibinfo  {journal} {Phys. Rev. B}\ }\textbf {\bibinfo {volume}
  {100}},\ \bibinfo {pages} {134407} (\bibinfo {year} {2019})}\BibitemShut
  {NoStop}%
\bibitem [{\citenamefont {El\'{\i}as}\ and\ \citenamefont
  {Verga}(2014)}]{QuantumFourier}%
  \BibitemOpen
  \bibfield  {author} {\bibinfo {author} {\bibfnamefont {R.~G.}\ \bibnamefont
  {El\'{\i}as}}\ and\ \bibinfo {author} {\bibfnamefont {A.~D.}\ \bibnamefont
  {Verga}},\ }\href {\doibase 10.1103/PhysRevB.89.134405} {\bibfield  {journal}
  {\bibinfo  {journal} {Phys. Rev. B}\ }\textbf {\bibinfo {volume} {89}},\
  \bibinfo {pages} {134405} (\bibinfo {year} {2014})}\BibitemShut {NoStop}%
\bibitem [{\citenamefont {Wieser}(2013)}]{Wieser1}%
  \BibitemOpen
  \bibfield  {author} {\bibinfo {author} {\bibfnamefont {R.}~\bibnamefont
  {Wieser}},\ }\href {\doibase 10.1103/PhysRevLett.110.147201} {\bibfield
  {journal} {\bibinfo  {journal} {Phys. Rev. Lett.}\ }\textbf {\bibinfo
  {volume} {110}},\ \bibinfo {pages} {147201} (\bibinfo {year}
  {2013})}\BibitemShut {NoStop}%
\bibitem [{\citenamefont {Wieser}(2015)}]{Wieser2}%
  \BibitemOpen
  \bibfield  {author} {\bibinfo {author} {\bibfnamefont {R.}~\bibnamefont
  {Wieser}},\ }\href {\doibase 10.1140/epjb/e2015-50832-0} {\bibfield
  {journal} {\bibinfo  {journal} {Eur. Phys. J. B}\ }\textbf {\bibinfo {volume}
  {88}} (\bibinfo {year} {2015}),\ 10.1140/epjb/e2015-50832-0}\BibitemShut
  {NoStop}%
\bibitem [{\citenamefont {Bogdanov}\ and\ \citenamefont
  {Yablonskii}(1989)}]{Bogdanov}%
  \BibitemOpen
  \bibfield  {author} {\bibinfo {author} {\bibfnamefont {A.~N.}\ \bibnamefont
  {Bogdanov}}\ and\ \bibinfo {author} {\bibfnamefont {D.~A.}\ \bibnamefont
  {Yablonskii}},\ }\href@noop {} {\bibfield  {journal} {\bibinfo  {journal}
  {Zh. Eksp. Teor. Fiz.}\ }\textbf {\bibinfo {volume} {95}},\ \bibinfo {pages}
  {178} (\bibinfo {year} {1989})}\BibitemShut {NoStop}%
\bibitem [{\citenamefont {Wang}\ \emph {et~al.}(2018)\citenamefont {Wang},
  \citenamefont {Yuan},\ and\ \citenamefont {Wang}}]{Size}%
  \BibitemOpen
  \bibfield  {author} {\bibinfo {author} {\bibfnamefont {X.}~\bibnamefont
  {Wang}}, \bibinfo {author} {\bibfnamefont {H.}~\bibnamefont {Yuan}}, \ and\
  \bibinfo {author} {\bibfnamefont {X.}~\bibnamefont {Wang}},\ }\href {\doibase
  10.1038/s42005-018-0029-0} {\bibfield  {journal} {\bibinfo  {journal} {Commun
  Phys}\ }\textbf {\bibinfo {volume} {1}} (\bibinfo {year} {2018}),\
  10.1038/s42005-018-0029-0}\BibitemShut {NoStop}%
\bibitem [{\citenamefont {Wilson}\ \emph {et~al.}(2014)\citenamefont {Wilson},
  \citenamefont {Butenko}, \citenamefont {Bogdanov},\ and\ \citenamefont
  {Monchesky}}]{Phases}%
  \BibitemOpen
  \bibfield  {author} {\bibinfo {author} {\bibfnamefont {M.~N.}\ \bibnamefont
  {Wilson}}, \bibinfo {author} {\bibfnamefont {A.~B.}\ \bibnamefont {Butenko}},
  \bibinfo {author} {\bibfnamefont {A.~N.}\ \bibnamefont {Bogdanov}}, \ and\
  \bibinfo {author} {\bibfnamefont {T.~L.}\ \bibnamefont {Monchesky}},\ }\href
  {\doibase 10.1103/PhysRevB.89.094411} {\bibfield  {journal} {\bibinfo
  {journal} {Phys. Rev. B}\ }\textbf {\bibinfo {volume} {89}},\ \bibinfo
  {pages} {094411} (\bibinfo {year} {2014})}\BibitemShut {NoStop}%
\bibitem [{\citenamefont {Petrovi\ifmmode~\acute{c}\else \'{c}\fi{}}\ \emph
  {et~al.}(2018)\citenamefont {Petrovi\ifmmode~\acute{c}\else \'{c}\fi{}},
  \citenamefont {Popescu}, \citenamefont {Bajpai}, \citenamefont
  {Plech\'a\ifmmode~\check{c}\else \v{c}\fi{}},\ and\ \citenamefont
  {Nikoli\ifmmode~\acute{c}\else \'{c}\fi{}}}]{Popescu}%
  \BibitemOpen
  \bibfield  {author} {\bibinfo {author} {\bibfnamefont {M.~D.}\ \bibnamefont
  {Petrovi\ifmmode~\acute{c}\else \'{c}\fi{}}}, \bibinfo {author}
  {\bibfnamefont {B.~S.}\ \bibnamefont {Popescu}}, \bibinfo {author}
  {\bibfnamefont {U.}~\bibnamefont {Bajpai}}, \bibinfo {author} {\bibfnamefont
  {P.}~\bibnamefont {Plech\'a\ifmmode~\check{c}\else \v{c}\fi{}}}, \ and\
  \bibinfo {author} {\bibfnamefont {B.~K.}\ \bibnamefont
  {Nikoli\ifmmode~\acute{c}\else \'{c}\fi{}}},\ }\href {\doibase
  10.1103/PhysRevApplied.10.054038} {\bibfield  {journal} {\bibinfo  {journal}
  {Phys. Rev. Applied}\ }\textbf {\bibinfo {volume} {10}},\ \bibinfo {pages}
  {054038} (\bibinfo {year} {2018})}\BibitemShut {NoStop}%
\bibitem [{\citenamefont {Mentink}\ \emph {et~al.}(2010)\citenamefont
  {Mentink}, \citenamefont {Tretyakov}, \citenamefont {Fasolino}, \citenamefont
  {Katsnelson},\ and\ \citenamefont {Rasing}}]{LLGnumerics}%
  \BibitemOpen
  \bibfield  {author} {\bibinfo {author} {\bibfnamefont {J.~H.}\ \bibnamefont
  {Mentink}}, \bibinfo {author} {\bibfnamefont {M.~V.}\ \bibnamefont
  {Tretyakov}}, \bibinfo {author} {\bibfnamefont {A.}~\bibnamefont {Fasolino}},
  \bibinfo {author} {\bibfnamefont {M.~I.}\ \bibnamefont {Katsnelson}}, \ and\
  \bibinfo {author} {\bibfnamefont {T.}~\bibnamefont {Rasing}},\ }\href
  {\doibase 10.1088/0953-8984/22/17/176001} {\bibfield  {journal} {\bibinfo
  {journal} {Journal of Physics: Condensed Matter}\ }\textbf {\bibinfo {volume}
  {22}},\ \bibinfo {pages} {176001} (\bibinfo {year} {2010})}\BibitemShut
  {NoStop}%
\bibitem [{\citenamefont {Dormand}\ and\ \citenamefont
  {Prince.}(1980)}]{ODE45}%
  \BibitemOpen
  \bibfield  {author} {\bibinfo {author} {\bibfnamefont {J.}~\bibnamefont
  {Dormand}}\ and\ \bibinfo {author} {\bibfnamefont {P.}~\bibnamefont
  {Prince.}},\ }\href@noop {} {\bibfield  {journal} {\bibinfo  {journal}
  {Journal of Computational and Applied Mathematics}\ }\textbf {\bibinfo
  {volume} {6}} (\bibinfo {year} {1980})}\BibitemShut {NoStop}%
\bibitem [{\citenamefont {Lent}\ and\ \citenamefont {Kirkner}(1990)}]{Lent}%
  \BibitemOpen
  \bibfield  {author} {\bibinfo {author} {\bibfnamefont {C.~S.}\ \bibnamefont
  {Lent}}\ and\ \bibinfo {author} {\bibfnamefont {D.~J.}\ \bibnamefont
  {Kirkner}},\ }\href {\doibase 10.1063/1.345156} {\bibfield  {journal}
  {\bibinfo  {journal} {Journal of Applied Physics}\ }\textbf {\bibinfo
  {volume} {67}},\ \bibinfo {pages} {6353} (\bibinfo {year}
  {1990})}\BibitemShut {NoStop}%
\bibitem [{\citenamefont {Osca}\ and\ \citenamefont {Serra}(2016)}]{Osca1}%
  \BibitemOpen
  \bibfield  {author} {\bibinfo {author} {\bibfnamefont {J.}~\bibnamefont
  {Osca}}\ and\ \bibinfo {author} {\bibfnamefont {L.}~\bibnamefont {Serra}},\
  }\href {\doibase 10.1140/epjb/e2016-70694-0} {\bibfield  {journal} {\bibinfo
  {journal} {Eur. Phys. J. B}\ }\textbf {\bibinfo {volume} {90}} (\bibinfo
  {year} {2016}),\ 10.1140/epjb/e2016-70694-0}\BibitemShut {NoStop}%
\bibitem [{\citenamefont {Osca}\ and\ \citenamefont {Serra}(2019)}]{Osca2}%
  \BibitemOpen
  \bibfield  {author} {\bibinfo {author} {\bibfnamefont {J.}~\bibnamefont
  {Osca}}\ and\ \bibinfo {author} {\bibfnamefont {L.}~\bibnamefont {Serra}},\
  }\href {\doibase 10.1140/epjb/e2019-100011-2} {\bibfield  {journal} {\bibinfo
   {journal} {The European Physical Journal B}\ }\textbf {\bibinfo {volume}
  {92}},\ \bibinfo {pages} {101} (\bibinfo {year} {2019})}\BibitemShut
  {NoStop}%
\bibitem [{Har()}]{Harwell}%
  \BibitemOpen
  \href@noop {} {}\bibinfo {note} {"HSL (2013). A collection of Fortran codes
  for large scale scientific computation. http://www.hsl.rl.ac.uk"}\BibitemShut
  {NoStop}%
\end{thebibliography}%

\end{document}